\documentstyle[pre,aps,eqsecnum,multicol]{revtex}
\begin{document}

\draft
\title{Instanton for the Kraichnan passive scalar problem}
\author{E. Balkovsky$^1$ and V. Lebedev$^{1,2}$}
\address{$^1$ Department of Physics of Complex Systems,
Weizmann Institute of Science, \\
Rehovot 76100, Israel; \\
$^2$ Landau Institute for Theoretical Physics, RAS, \\
Kosygina 2, Moscow, 117940, Russia.}
%\date{\today}
\maketitle

\begin{abstract}
We consider high-order correlation functions of the passive scalar in the
Kraichnan model. Using the instanton formalism we find the scaling exponents
$\zeta_n$ of the structure functions $S_n$ for $n\gg1$ under the additional
condition ${d}\zeta_2\gg1$ (where ${d}$ is the dimensionality of
space). At $n<n_c$ (where $n_c={d}\zeta_2/[2(2-\zeta_2)]$) the exponents
are $\zeta_n=(\zeta_2/4)(2n-n^2/n_c)$, while at $n>n_c$ they are
$n$ independent: $\zeta_n=\zeta_2 n_c/4$.  We also estimate $n$-dependent
factors in $S_n$, particularly their behavior at $n$ close to $n_c$. 
\end{abstract}


\begin{multicols}{2}

\section*{Introduction}

Anomalous scaling is probably the central problem of the theory of turbulence.
In 1941 Kolmogorov formulated his famous theory of developed turbulence
\cite{Kol}, where the scaling behavior of different correlation functions of the
turbulent velocity was predicted.  Experimentally one observes deviations from
the scaling exponents, proposed by Kolmogorov \cite{61GSM,62BT,91MS}. It is
recognized that the deviations are related to rare strong fluctuations making
the main contribution into the correlation functions \cite{53Bat,MY,Frish}.
This phenomenon, which is usually called intermittency, is the most striking
peculiarity of developed turbulence.

One of the classical objects in the theory of turbulence is a passive scalar
advected by a fluid. The role of the passive scalar can be played by
temperature or by the density of pollutants. Correlation functions of the
scalar in a turbulent flow possess a scaling behavior that was established by
Obukhov and Corrsin \cite{Obukh,Corrs} in the frame of the theory analogous to
that of Kolmogorov . Intermittency enforces deviations from the Obukhov-Corrsin
exponents that appear to be even stronger than the deviations from the
Kolmogorov exponents for the correlation functions of the velocity
\cite{84AHGA,90MS,91Sre,91HY}.

Unfortunately, a consistent theory of turbulence describing anomalous scaling
has not been constructed yet. This accounts for the difficulties associated with
the strong coupling inherent to developed turbulence. This is the reason for
attempts to examine the intermittency phenomenon in the framework of different
simplified models. The most popular model used for this purpose is
Kraichnan's model of passive scalar advection \cite{68Kra-a}, where the
advecting velocity is believed to be short correlated in time and have
Gaussian distribution. That allows one to examine the statistics of the 
passive scalar in more detail. 

The scalar in the Kraichnan model exhibits strong intermittency even if it is
absent in the advecting velocity field. This was proved both theoretically
\cite{95SS,95GK,95CFKLb,95CF,96BGK,96SS,96BG,97BFL} and numerically
\cite{95KYC,97FGLP,98FMV}. In the theoretical works the equation for the
$n$-point correlation function $F_n$ was solved assuming that different
parameters, such as $\zeta_2$, $2-\zeta_2$, or ${d}^{-1}$, are small (recall,
that $\zeta_2$ is the exponent of the second-order correlation function of the
passive scalar and ${d}$ is the dimensionality of space). The order of the
correlation functions that can be examined in the framework of the methods of
the noted papers is bounded from above, which does not allow one to imagine the
whole dependence of $\zeta_n$ on $n$. For that it would be enough to get the
asymptotic behavior of $\zeta_n$ at $n\gg 1$. There have been several attempts
to find the scaling of the correlation functions for larger $n$. In the the
work by Kraichnan \cite{closure} a closure was assumed enabling one to find
$\zeta_n$ for any $n$. An alternative scheme was proposed in \cite{97Yak}. An
attempt to solve the problem at large $n$ was made in \cite{97Chert}, where an
$n$-independent asymptotic behavior was found.

In the present work we develop a technique based on the path-integral
representation of the dynamical correlation functions of classical fields
\cite{73MSR,76Dom,76Jan}. We use an idea, formulated in \cite{96FKLM}, that is
related to the possibility of exploiting the saddle-point approximation in the
path integral at large $n$. The saddle-point conditions are
integro-differential equations describing an object that, in analogy to the
quantum field theory, we call an instanton. The instantonic method was already
successfully used in some contexts. Results concerning Burgers turbulence,
conventional Navier-Stokes turbulence, and modifications of the Kraichnan
model were obtained with the help of this method in the Refs.
\cite{96GM,97BFKL,97FL,98BF}. The formalism presented in this paper enables one
to find correlation functions of the passive scalar for arbitrary $n\gg1$
provided  ${d}\zeta_2\gg1$.

The paper is organized as follows. In Sec. \ref{model} we formulate the
Kraichnan model, introduce notation, and write down the standard path integral
representation for the correlation functions. This basic representation turns
out to be  unsuitable for the saddle-point approximation; therefore, we
reformulate the problem in Sec. \ref{lagrange}. Passing to new variables that
are Lagrangian separations, we get a path integral that already admits the use
of the saddle-point approximation. In Sec. \ref{instant} we consider the
instantonic equations for the case of the structure functions. We
solve these equations in the limit ${d}\zeta_2\gg1$, which enables us to
find the anomalous scaling and estimate the $n$ dependence of $S_n$. The main
results of the work are presented in Sec. \ref{results} and discussed in
Conclusion. Details of calculations are given in Appendixes.

\section{Kraichnan Model}
\label{model}

Advection of a passive scalar $\theta$ by a velocity field
${\bbox v}$ is described by the equation
\begin{equation}
\partial_t\theta+{\bbox v}\nabla\theta
-\kappa\nabla^2\theta=\phi \,,
\label{adv} \end{equation}
where $\kappa$ is the diffusion coefficient and $\phi$ is the source of the
passive scalar (say, if $\theta$ corresponds to fluctuations of temperature,
then $\phi$ represents the power of heaters). In a turbulent flow, ${\bbox v}$
is a random function of time and space coordinates. The source $\phi$ is also
assumed to be a random function. Then passive scalar correlation functions are
determined by the statistics of ${\bbox v}$ and $\phi$.  Usually, one is
interested in simultaneous correlation functions
$F_n=\left\langle\theta({\bbox r}_1)\cdots\theta({\bbox r}_n)\right\rangle$
since a large-scale velocity
destroys temporal correlations in the Eulerian frame, whereas simultaneous
objects are not influenced by it.

It is convenient to examine the anomalous scaling in terms of the
structure functions
\begin{equation}
S_n(r)=\left\langle|\theta({\bbox r}/2)
-\theta(-{\bbox r}/2)|^{n}\right\rangle \,.
\label{str} \end{equation}
One expects a universal behavior of the structure functions in the
convective interval of scales $r_d\ll r \ll L$, where $r_d$ is the
scale where the diffusivity becomes relevant and $L$ is the correlation
length of the scalar source $\phi$. Namely, one observes a scaling dependence
on $r$:
\begin{equation}
S_n(r)\propto r^{\zeta_n} \,.
\label{zeta} \end{equation}
In the frame of the Obukhov theory
\cite{Obukh,Corrs} $\zeta_n=(n/2)\zeta_2$. Therefore, the differences
$(n/2)\zeta_2-\zeta_n$, which are usually called anomalous exponents,
characterize the anomalous scaling. One can write an
estimate
\begin{equation}
S_n(r)\sim A_n\left[S_2(r)\right]^{n/2}\left(\frac{L}{r}\right)
^{(n/2)\zeta_2-\zeta_n} \,,
\label{inter} \end{equation}
where $A_n$ is an $n$-dependent factor. Note that
Eq. (\ref{inter}) implies that the structure functions in the convective
interval do not depend on the diffusion length $r_d$. The intermittency leads to
the conclusion that values of the structure functions should be much larger
than their naive Obukhov estimations \cite{Frish}. Therefore
$(n/2)\zeta_2-\zeta_n>0$ and we conclude that these are the anomalous exponents
that reflect the intermittency.  

\subsection{Formulation of the problem}

In the Kraichnan model both ${\bbox v}$ and $\phi$ are assumed to be 
independent random functions, $\delta$ correlated in time and described by 
Gaussian statistics homogeneous in space. Therefore, statistical properties
of the fields are entirely characterized by the pair correlation functions
\begin{eqnarray} &&
\langle\phi(t_1,{\bbox r}_1)
\phi(t_2,{\bbox r}_2)\rangle
=\chi(r_{12})
\delta(t_1-t_2) \,,
\quad \chi(0)=P_2 \,,
\nonumber\\ &&
\langle v_\alpha(t_1,{\bbox r}_1)v_\beta(t_2,{\bbox r}_2)\rangle
={\cal V}_{\alpha\beta}({\bbox r}_1-{\bbox r}_2)\delta(t_1-t_2)\,.
\nonumber\end{eqnarray}
Here, $\chi(r)$ is a smooth function decaying on the scale $L$ which is the
pumping length. The constant $P_2$ has the meaning of the pumping rate of
$\theta^2$. The tensor ${\cal V}_{\alpha\beta}(r)$ has a characteristic
scale $L_v$, which has the meaning of the pumping length of the velocity.
We will assume that $L_v\gg L$. Since $r\ll L$ in the convective interval,
we will need ${\cal V}_{\alpha\beta}$ only at $r\ll L_v$, where one can write
\begin{equation}
{\cal V}_{\alpha\beta}({\bbox r})={\cal V}_0\delta_{\alpha\beta}
-{\cal K}_{\alpha\beta}({\bbox r})\,.
\label{bl2}\end{equation}
The quantity ${\cal V}_0$ is an $r$-independent constant that is the main
contribution to the velocity correlation function on scales less than the
velocity pumping length $L_v$. Nevertheless, besides ${\cal V}_0$, we should
also keep a small $r$-dependent correction ${\cal K}$, since ${\cal V}_0$
corresponds to advection homogeneous in space and therefore 
does not contribute to simultaneous correlation functions of $\theta$.

The velocity correlation function is assumed to possess some scaling
properties. Namely, ${\cal K}({\bbox r})\propto r^{2-\gamma}$, where the
exponent $\gamma$ characterizes the roughening degree of the velocity field. 
The field is smooth in space at $\gamma=0$ and is extremely irregular at
$\gamma=2$.  We will treat an arbitrary $\gamma$ satisfying the inequality
$0<\gamma<2$. The tensorial structure of ${\cal K}_{\alpha\beta}$ is determined
by the incompressibility condition ${\rm div}\,{\bbox v}=0$, implied
in the Kraichnan model
\begin{equation}
{\cal K}_{\alpha\beta}({\bbox r})=\frac{D}{d}
r^{-\gamma}\left[\frac{2-\gamma}{{d}-1}
(r^2\delta_{\alpha\beta}-r_\alpha r_\beta)
+r^2\delta_{\alpha\beta}\right] \,.
\label{eq2} \end{equation}
Here $d$ is the dimensionality of space and $D$ is a constant
characterizing the strength of velocity fluctuations.   
One assumes that the fluctuations are strong enough to ensure
the large value of the Peclet number, that is,
\begin{equation}
DL^{2-\gamma}\gg\kappa \,.
\label{mo3} \end{equation}
The inequality (\ref{mo3}) ensures the existence of the convective interval
of scales since it can be rewritten as $r_d\ll L$, where $r_d$ is the 
diffusive length
\begin{equation}
r_d^{2-\gamma}\sim\kappa/D \,.
\label{mo2} \end{equation}

The assumption of the Gaussian nature and zero correlation time for the fields 
${\bbox v}$ and $\phi$ allows one to derive a closed partial differential
equation for the $n$th-order correlation function $F_n$ of $\theta$
\cite{68Kra-a,94SS,95CFKLb}. For the simultaneous pair correlation function
$F_2(r_{12})=\langle\theta(t,{\bbox r}_1)\theta(t,{\bbox r}_2)\rangle$ 
one can solve the equation and find the explicit expression for $F_2$.
In the convective interval \cite{68Kra-a}
\begin{equation}
S_2(r)=2[F_2(0)-F_2(r)]\sim \frac{P_2}{D} r^\gamma \,. 
\label{pair} \end{equation}
Comparing Eq. (\ref{pair}) with Eq. (\ref{zeta}), one concludes that the
exponent  $\gamma$ introduced by Eq. (\ref{eq2}) directly determines the
scaling of the second-order structure function $\zeta_2=\gamma$.

However, for $n>2$ the equations for $F_n$ are too complicated to be 
integrated exactly. In \cite{95GK,95CFKLb,95CF,96BGK} the equations were 
analyzed in the limits $2-\gamma\ll1$ and ${d}\gamma\gg 1$, where
the statistics of the passive scalar is close to Gaussian. The analysis
led to an anomalous scaling that can be expressed in terms of the 
exponents $\zeta_n$ of the structure functions (\ref{str}) and (\ref{zeta})
\begin{equation}
\zeta_n=\frac{n\gamma}{2}-\frac{2-\gamma}{2({d}+2)}n(n-2) \,.
\label{pert} \end{equation}
This expression covers both limit cases $2-\gamma\ll1$ and 
${d}\gamma\gg1$. The first term on the right-hand side
of Eq. (\ref{pert}) represents the normal scaling whereas the second 
one is just the anomalous scaling exponent. The calculations 
leading to Eq. (\ref{pert}) are correct if the anomalous contribution 
is much smaller than the normal one, which implies the inequality
\begin{equation}
n\ll \frac{{d}\gamma}{2-\gamma} \,. 
\label{pert1} \end{equation}
Below we will develop a different approach to the problem. It will allow us
to find the exponents $\zeta_n$ [Eq. (\ref{zeta})] of the structure correlation 
functions (\ref{str}) for any order $n \gg 1$ under the same additional
condition ${d}\gamma\gg1$ as in \cite{95CFKLb,95CF}.

\subsection{Path integral}

Generally, the statistics of classical fields in the presence of random
forces can be examined with the help of the field technique formulated in
\cite{73MSR,76Dom,76Jan}. In the framework of the technique, correlation
functions are calculated as path integrals with the weight $\exp(i{\cal I})$,
where ${\cal I}$ is the effective action related to dynamical equations for the
fields. For the passive scalar in the Kraichnan model the effective action is
\begin{eqnarray} &&
i{\cal I}_\theta=i\int {\rm d}t\,{\rm d}{\bbox r}\,
\left[p\partial_t\theta+p{\bbox v\nabla}\theta
+\kappa\nabla p\nabla\theta\right]
\nonumber \\ &&
-\frac{1}{2}\int {\rm d}t\,{\rm d}{\bbox r}_1\,{\rm d}{\bbox r}_2\,
\chi\left(|{\bbox r}_1-{\bbox r}_2|\right)
p(t,{\bbox r}_1)p(t,{\bbox r}_2)\,,
\label{bl1} \end{eqnarray}
where $p$ is an auxiliary field conjugated to $\theta$.
The first term in the effective action (\ref{bl1}) is directly related to the
left-hand side of Eq. (\ref{adv}). The quadratic in the $p$ term in Eq.
(\ref{bl1})
appears as a result of averaging over the statistics of the pumping $\phi$.

Simultaneous correlation functions of $\theta$ can be represented
as functional derivatives of the generating functional 
\begin{equation} {\cal Z}(\lambda)=\left\langle\exp\left[i\int {\rm d}{\bbox r} 
\lambda({\bbox r})\theta(t=0,{\bbox r})\right]\right\rangle \,, 
\label{zlam}\end{equation}
where angular brackets designate averaging over the statistics of $\phi$ 
and ${\bbox v}$. With the help of the action (\ref{bl1}) the generating 
functional can be rewritten as the path integral
\begin{eqnarray} &&
{\cal Z}(\lambda)=
\int{\cal D}\theta\,{\cal D}p\,{\cal D}{\bbox v}\,
\exp\biggl[-{\cal F}({\bbox v}) 
\nonumber \\ &&  
+i{\cal I}_\theta+i\int {\rm d}{\bbox r}\,
\lambda({\bbox r})\theta(t=0,{\bbox r})\biggr] \,.
\label{bl3} \end{eqnarray}
Here ${\cal F}({\bbox v})$ determines the statistics of the velocity field. 
Since we assume the Gaussian nature of the statistics, ${\cal F}({\bbox v})$ is
a functional of second order over ${\bbox v}$ with the kernel determined by
the pair correlation function (\ref{bl2}). Knowing ${\cal Z}(\lambda)$
one can restore the probability distribution function (PDF) of $\theta$. It is
convenient to treat the PDF of a particular object
\begin{equation}
\vartheta=\int {\rm d}{\bbox r}\,
\beta({\bbox r})\theta(t=0,{\bbox r}) \,,
\label{bb14} \end{equation}
with a given function $\beta({\bbox r})$. For example, the set of
the structure functions (\ref{str}) can be assembled into the PDF of
the scalar difference in two points 
$\theta({\bbox r}/2)-\theta(-{\bbox r}/2)$
which is the object (\ref{bb14}) with 
$\beta({\bbox r}_1)=\delta({\bbox r}_1-{\bbox r}/2)
-\delta({\bbox r}_1+{\bbox r}/2)$. The PDF of $\vartheta$ is written as
\begin{equation}
{\cal P}(\vartheta)=
\int_{-\infty}^{\infty}\frac{{\rm d} y}{2\pi}\exp(-iy\vartheta)
{\cal Z}[y\beta({\bbox r})] \,.
\label{bb15} \end{equation}
Moments of $\vartheta$ are then expressed as
\begin{equation}
\langle|\vartheta|^n\rangle
=\int_{-\infty}^{\infty} {\rm d}\vartheta\,|\vartheta|^n
{\cal P}(\vartheta) \,.
\label{mo1} \end{equation}

We will be interested in the high-order correlation functions of $\vartheta$
or, in other words, we consider the limit $n\gg 1$. This is equivalent to
examining the large $\vartheta$ tail of the PDF (\ref{bb15}). One could
expect \cite{96FKLM} that the tail can be calculated in the saddle-point
approximation since there is a large parameter $\vartheta$ in the
corresponding path integral. Unfortunately, direct application of the method to
the integral (\ref{bb15}) or to the moments (\ref{mo1}) does not lead to
success.  

To recognize the reason, let us consider the
transformation of the variables \cite{96FKLM} (see also \cite{97FL})
\begin{eqnarray} &&
{\bbox v}\to X {\bbox v}\,,\ 
p\to Xp\,,\ t\to X^{-1}t\,,\
y\to Xy\,,\ \kappa\to X\kappa \,.
\nonumber \end{eqnarray}
One can check that under this transformation all the terms in the square
brackets in the right-hand side of Eq. (\ref{bl3}) acquire the factor $X$, which
means that in the saddle-point approximation $\ln\left[{\cal
Z}(y\beta)\right]=y f(y/\kappa)$ with some unknown function $f$. On the other
hand, we expect that correlation functions of the scalar itself (but not of its
gradient, for example) do not depend on the diffusivity and the results of the
works \cite{95SS,95GK,95CFKLb,95CF,96BGK,96BG,97BFL,94SS} confirm the
expectation. Then, at small $\kappa$ the function $f$ remains a
$\kappa$-independent constant and we obtain
\begin{equation}
\ln{\cal Z}(y\beta)\propto|y| \,.
\label{tr3} \end{equation}
Unfortunately, Eq. (\ref{tr3}) does not help to restore ${\cal P}(\vartheta)$
since after substituting it into Eq. (\ref{bb15}) we realize that the
characteristic value of $y$ in the integral can be estimated as
$y\sim\vartheta^{-1}$. Therefore, at large $\vartheta$ the main contribution to
the integral is determined by the region where Eq. (\ref{tr3}) does not work.

We conclude that the naive instantonic approach to the problem fails. The
reason is that for the instanton the velocity field is fixed (does not
fluctuate) in time and space. Obviously, a saddle-point solution is anisotropic
because of the incompressibility condition ${\rm div}\,{\bbox v}=0$. 
Fluctuations related to smooth variations of the anisotropy axis in time and
space are strong and destroy the saddle-point approximation for the tail of the
PDF ${\cal P}(\vartheta)$ or for the high moments of $\vartheta$. Thus we
should transform the problem to more adequate variables, where fluctuations of
the velocity are partly taken into account. This is the only chance to
construct an instanton with weak fluctuations on its background. This is the
goal of the next section.

\section{Lagrange Formulation}
\label{lagrange}

As we mentioned above, the diffusivity $\kappa$ does not enter the result for
the structure functions. Therefore, we will assume $\kappa=0$ in all the
following calculations. However, one should be careful since in this case it is
impossible to deal with point objects. To provide a regularization, we should
assume that the characteristic scales of the function $\beta$ in Eq. 
(\ref{bb14}) are larger than $r_d$. In addition, the scales are to be much
smaller than $L$ since we are going to examine correlation functions in the
convective interval.

In the diffusionless case the left-hand side of Eq. (\ref{adv}) describes the
field $\theta$ moving together with the fluid. Then it is natural to pass into
the Lagrangian frame where the process is trivial. For that purpose we
introduce Lagrangian trajectories ${\bbox\varrho}(t)$ that obey the equation
\begin{equation}
\partial_t{\bbox\varrho}={\bbox v}(t,{\bbox\varrho}) \,.
\label{bl4} \end{equation}
We will label the trajectories by the positions of fluid particles at $t=0$: 
${\bbox\varrho}(t=0)={\bbox r}$. Equation (\ref{adv})
(where $\kappa$ is omitted) 
can easily be solved in terms of the Lagrangian trajectories 
\begin{eqnarray} && 
\theta(0,{\bbox r})=\int_{-\infty}^0\!{\rm d}t\,
\phi\left[t,{\bbox\varrho}(t,{\bbox r})\right] \,,  
\label{grf}\end{eqnarray}
Since we are interested in the field $\theta$ at $t=0$, due to causality
the integration is performed over negative time. Therefore, Eq. (\ref{bl4})
should be solved backwards in time. 

A simultaneous $n$th-order correlation function of $\theta$ can be written as
the product of $n$ integrals (\ref{grf}), averaged over the statistics of
${\bbox v}$ and $\phi$. In this representation, averaging over the pumping
is very simple. For example, the two-point correlation function is
\begin{eqnarray} &&
F_2=\int_{-\infty}^0\!{\rm d}t\,
\left\langle\chi\left(R_{12}\right)\right\rangle_v \,,
\label{F2} \\ && 
{R}_{12}(t)\equiv{R}(t,{\bbox r}_1,{\bbox r}_2)=
|{\bbox\varrho}(t,{\bbox r}_1)
-{\bbox\varrho}(t,{\bbox r}_2)| \,.
\label{la30}\end{eqnarray}
The angular brackets $\langle\rangle_v$ in Eq. (\ref{F2}) denote
averaging over the statistics of ${\bbox v}$ only, since the statistics
of $\phi$ is already
accounted for there. Similar formulas can be written for correlation
functions
of higher orders. Once this is done, one can assemble them into the generating
functional (\ref{zlam})
\begin{eqnarray}
{\cal Z}(\lambda)=\left\langle\exp\left\{-
\frac{1}{2}\int {\rm d}t{\rm d}{\bbox r}_{1}{\rm d}{\bbox r}_{2}\,
\chi\left(R_{12}\right)
\lambda_1\lambda_2\right\}\right\rangle_v  \,.
\label{zlv} \end{eqnarray}
where $\lambda_{1,2}=\lambda({\bbox r}_{1,2})$.
Calculating the moments of the object (\ref{bb14}) 
in accordance with Eqs. (\ref{bb15}) and (\ref{mo1}) we get
\begin{eqnarray} &&
\langle|\vartheta|^n\rangle=\int\frac{{\rm d}y\,{\rm d}\vartheta}{2\pi}
\left\langle\exp\left(-{\cal F}_\lambda
-iy\vartheta +n\ln|\vartheta|\right)\right\rangle_v \,,
\label{bb86} \\ &&
{\cal F}_\lambda=
\frac{y^2}{2}\int {\rm d}t\,{\rm d}{\bbox r}_1\,{\rm d}{\bbox r}_2\,
\chi(R_{12})\beta({\bbox r}_1)\beta({\bbox r}_2) \,.
\label{bb76} \end{eqnarray}
At this point, we would like to stress the close connection between the
statistics of the passive scalar and that of Lagrangian trajectories
\cite{Les}, which can be seen from Eq. (\ref{zlv}).

\subsection{Statistics of Lagrangian separations}

Equations (\ref{zlv},\ref{bb86}) show that
the correlation functions we are interested in
us are expressed via the average of $\exp(-{\cal F}_\lambda)$ over the
velocity. Note that ${\cal F}_\lambda$ [Eq. (\ref{bb76})] depends only on the
absolute values $R_{12}(t)$ of Lagrangian differences (\ref{la30}).  Therefore,
instead of averaging over the statistics of ${\bbox v}$ one could find the
answer by averaging over the statistics of the Lagrangian separations $R_{12}$.
Due to zero correlation time of the velocity field, the statistical properties
of the field $R_{12}$ appear to be relatively simple.

To establish the statistics of $R_{12}$ we start from the relation
\begin{equation} 
\gamma^{-1}\partial_t
R_{12}^{\gamma}=\zeta_{12}\equiv 
R_{12}^{\gamma-2}R_{12\alpha}
(v_{1\alpha}-v_{2\alpha}) \,,  
\label{la31} \end{equation} 
following from Eqs. (\ref{bl4}) and (\ref{la30}). As shown in
Appendix \ref{rich}, the average value of $\zeta_{12}$ is nonzero: 
\begin{equation}
\langle\zeta_{12}\rangle=-D \,.  
\label{aa35} \end{equation} 
Next, exploiting the expression (\ref{bl2}) for the velocity correlation
function, one can find the irreducible pair correlation function 
\begin{equation}
\langle\langle\zeta_{12}(t_1)\zeta_{34}(t_2)\rangle\rangle 
=\frac{2D}{{d}}Q_{12,34}\delta(t_1-t_2)\,.  
\label{la33} \end{equation} 
The explicit expression for the function $Q$ is rather cumbersome
\end{multicols}
\begin{eqnarray} &&
Q_{12,34}=\frac{{d}+1-\gamma}{4({d}-1)}
R_{12}^{\gamma-2}R_{34}^{\gamma-2} \left(R_{23}^{2-\gamma}+R_{14}^{2-\gamma}
-R_{13}^{2-\gamma}-R_{24}^{2-\gamma}\right)
\left(R_{23}^{2}+R_{14}^{2}-R_{13}^{2}-R_{24}^{2}\right) 
\nonumber \\ &&
-\frac{2-\gamma}{8({d}-1)} R_{12}^{\gamma-2}R_{34}^{\gamma-2}\biggl\{
\frac{1}{R_{13}^\gamma} \left(R_{12}^{2}+R_{13}^{2}-R_{23}^{2}\right)
\left(R_{13}^{2}+R_{34}^{2}-R_{14}^{2}\right)
+\frac{1}{R_{23}^\gamma} \left(R_{12}^{2}+R_{23}^{2}-R_{13}^{2}\right)
\left(R_{34}^{2}+R_{23}^{2}-R_{24}^{2}\right) 
\nonumber \\ &&
+\frac{1}{R_{14}^\gamma}
\left(R_{12}^{2}+R_{14}^{2}-R_{24}^{2}\right)
\left(R_{14}^{2}+R_{34}^{2}-R_{13}^{2}\right) 
+\frac{1}{R_{24}^\gamma} \left(R_{12}^{2}+R_{24}^{2}-R_{14}^{2}\right)
\left(R_{34}^{2}+R_{24}^{2}-R_{23}^{2}\right) \biggr\} \,. 
\label{la34} \end{eqnarray}
\begin{multicols}{2}

\noindent
It can be found from the definition of $\zeta_{12}$ [Eq. (\ref{la31})],
formula (\ref{eq2}) and relations such as 
\begin{eqnarray}&&
{\bbox R}_{12}{\bbox R}_{13} 
=\frac{1}{2}(R_{12}^2+R_{13}^2-R_{23}^2) \,,
\nonumber \\ &&
{\bbox R}_{12}{\bbox R}_{34} 
=\frac{1}{2}(R_{14}^2+R_{23}^2-R_{13}^2-R_{24}^2)\,.  
\nonumber \end{eqnarray}

In the spirit of the conventional procedure \cite{73MSR,76Dom,76Jan}, one can
assert that any average over the statistics of $R_{12}$ can be found as the
path integral over $R_{12}$ and over an auxiliary field 
$m_{12}\equiv m(t,{\bbox r}_1,{\bbox r}_2)$ with the weight
\begin{eqnarray}
\left\langle\exp\left[i \int {\rm d}t\,
{\rm d}{\bbox r}_1\,{\rm d}{\bbox r}_2\, 
\left(m_{12}\gamma^{-1}\partial_t R_{12}^\gamma
-m_{12}\zeta_{12}\right)\right]\right\rangle_v \,,
\nonumber \end{eqnarray}
where angular brackets mean averaging over the statistics of the velocity. 
Since $\zeta_{12}$ is $\delta$ correlated in time, the average can be expressed in
terms of Eqs. (\ref{aa35}) and (\ref{la33}) only. The result is 
$\exp\left(i{\cal I}_R\right)$, where
\begin{eqnarray} &&
i{\cal I}_R=i \int\limits_{-\infty}^0 {\rm d}t\,
\int {\rm d}{\bbox r}_1\,{\rm d}{\bbox r}_2\,
m_{12}(\gamma^{-1}\partial_t R_{12}^\gamma+D)
\nonumber \\ &&
-\frac{D}{{d}}\int\limits_{-\infty}^0 {\rm d}t\,
\int {\rm d}{\bbox r}_1\,{\rm d}{\bbox r}_2\,
{\rm d}{\bbox r}_3\,{\rm d}{\bbox r}_4\,Q_{12,34}m_{12}m_{34} \,.
\label{la36} \end{eqnarray}
Now, instead of Eq. (\ref{bb86}) we can write
\begin{eqnarray} &&
\langle|\vartheta|^n\rangle=\int\frac{{\rm d}y\,{\rm d}\vartheta}{2\pi}
\int {\cal D}R\,{\cal D}m\,
e^{i{\cal I}_R-{\cal F}_\lambda
-iy\vartheta+n\ln|\vartheta|} \,.
\label{bb57} \end{eqnarray}

The integration in Eq. (\ref{bb57}) is performed over functions of $t$, ${\bbox
r}_1$, and ${\bbox r}_2$ with some boundary conditions imposed on them. The
condition for the field $R_{12}$ follows from ${\bbox\varrho}(0)={\bbox r}$
and reads
\begin{equation}
R_{12}(t=0)=|{\bbox r}_1-{\bbox r}_2| \,.
\label{term} \end{equation} 
The boundary condition for the field  $m_{12}$ should be $m_{12}(-\infty)=0$, 
since we deal with free integration over $R_{12}$ in the remote past. Note 
that due to the definition (\ref{la30}) the triangle inequalities 
\begin{equation}
R_{12}+R_{23}>R_{13} 
\label{trian} \end{equation}
should be satisfied for any three points. Actually, the inequalities are
constraints that should be imposed on the field $R_{12}$ when
integrating in Eq. (\ref{bb57}). 

\subsection{General instantonic equations}
\label{genin}

In the preceding subsection we derive a formula (\ref{bb57}) for
$\langle|\vartheta|^n\rangle$. Its calculation is equivalent to solving
some nonlinear field theory. It looks very infeasible to perform this task.
We are going to calculate the integral (\ref{bb57}) in the saddle-point
approximation regarding the number $n$ large enough. To be consistent, when
doing the procedure one should remember about the constraints (\ref{trian}).
Unfortunately, it is very hard to take them into account explicitly.  We will
ignore the constraints, which is correct under the following conditions.
First, the inequalities (\ref{trian}) should be valid in the instantonic
solution. Second, fluctuations on the background of the instanton should be
weak (this is also the applicability condition of the instantonic formalism
itself). We argue in Appendixes \ref{triangle} and \ref{fluctu} that those
conditions are satisfied if
\begin{equation}
{d}\gamma\gg1 \,. 
\label{ineq} \end{equation}
Note also that for the condition (\ref{ineq}) fluctuations of a Lagrangian
separation near its average value are weak (see Appendix \ref{simul}).
The inequality (\ref{ineq}) will be implied below.

Thus we obtain from the integral (\ref{bb57}) in the saddle-point approximation
\begin{equation}
\langle|\vartheta|^n\rangle\sim
\left.\exp\left(i{\cal I}_R-{\cal F}_\lambda-iy\vartheta
+n\ln|\vartheta|\right)\right|_{\rm inst} \,.
\label{as20} \end{equation}
Here solutions of the instantonic equations should be substituted which are
extrema conditions for the argument of the exponent on the right-hand side of
Eq. (\ref{bb57}). Variation over $m_{12}$ and $R_{12}$ gives the
instantonic equations
\begin{eqnarray} &&
i(\gamma^{-1}\partial_t R_{12}^\gamma+D)
=2\frac{D}{{d}}\int {\rm d}{\bbox r}_3\,
{\rm d}{\bbox r}_4\,Q_{12,34}m_{34}\,,
\label{la37} \\ &&
iR_{12}^{\gamma-1}\partial_tm_{12}
\!+\!\frac{D}{{d}}\int {\rm d}{\bbox r}_3\,{\rm d}{\bbox r}_4\,
\left\{2\frac{\partial Q_{12,34}}{\partial R_{12}}m_{12}m_{34}\right.
\nonumber \\ &&
\left.+4\frac{\partial Q_{13,24}}{\partial R_{12}}m_{13}m_{24}\right\}
=-\frac{y^2}{2}\chi'(R_{12})\beta({\bbox r}_1)\beta({\bbox r}_2) \,.
\label{la38} \end{eqnarray}
The extremum conditions over $y$ and $\vartheta$ read
\begin{eqnarray} &&
\vartheta=iy\int {\rm d}t\,{\rm d}{\bbox r}_1\,{\rm d}{\bbox r}_2\,\chi(R_{12})
\beta({\bbox r}_1)\beta({\bbox r}_2)\,, 
\label{aa38} \\&&
iy=n/\vartheta \,.
\label{qq2} \end{eqnarray}
Note that only Eqs. (\ref{la37}) and (\ref{la38}) are true dynamical equations,
carrying the information about the dynamics of the flow, whereas Eqs. 
(\ref{aa38}) and (\ref{qq2}) are constraints imposed on the instantonic
solution. One needs to add to Eqs. (\ref{la37}) and (\ref{la38}) some boundary
conditions. The value of the field $R_{12}$ is fixed at $t=0$ by Eq.
(\ref{term}). As for the field $m_{12}$, we already noted that it should tend to
zero when $t\to-\infty$. It can be understood as the extremum condition that
appears after variation of the effective action over the boundary value of
$R_{12}$ in the remote past.

One can easily establish the asymptotic behavior of $R_{12}$
at $|t|\to\infty$. There the field $R_{12}$ grows and loses its dependence on
${\bbox r}_{1,2}$. The field $m_{12}$ tends to its vacuum zero
value at $|t|\to\infty$. Therefore, at large $|t|$ the term with $m_{12}$ in
Eq. (\ref{la37}) can be omitted and we find 
\begin{equation}
R^\gamma\approx\gamma D|t| \,.  
\label{aa40} \end{equation} 
The expression (\ref{aa40}) is nothing but the Richardson law for divergence
of Lagrangian trajectories \cite{26Rich}. Let us stress that now it
holds on the classical (mean-field) level, without taking into account
fluctuations on the background of the instanton. To clarify this point, notice
that if the velocity field is a deterministic function of time and space (as
it is for the naive instanton discussed above) then the Richardson law can
not be valid for all the Lagrangian trajectories. In our instanton we get rid
of the velocity field that resulted in the emergence of the Richardson law. 
Note that the triangle inequalities (\ref{trian}) are obviously satisfied both
for (\ref{term}) and for the asymptotic behavior (\ref{aa40}). 

The expression for the action appearing in Eq. (\ref{bb57}) is
\begin{eqnarray}&&
i{\cal I}\equiv i{\cal I}_R\!-\!{\cal F}_\lambda
\!=\!i\int {\rm d}t{\rm d}{\bbox r}_1{\rm d}{\bbox r}_2\,
\gamma^{-1}m_{12}\partial_t R_{12}^\gamma \!-\! E \,,
\label{aa60} \\ &&
E=\frac{y^2}{2}\!\int {\rm d}{\bbox r}_1{\rm d}{\bbox r}_2
\chi(R_{12})\beta({\bbox r}_1)\beta({\bbox r}_2)
-iD\int {\rm d}{\bbox r}_1{\rm d}{\bbox r}_2 m_{12}
\nonumber \\ &&
+\frac{D}{{d}}\int {\rm d}{\bbox r}_1\,{\rm d}{\bbox r}_2\,
{\rm d}{\bbox r}_3\,{\rm d}{\bbox r}_4\,Q_{12,34}m_{12}m_{34} \,.
\label{la39} \end{eqnarray}
We see from Eq. (\ref{aa60}) that the quantity $E$ plays the role of the
Hamiltonian function of the system, while Eqs. (\ref{la37}) and (\ref{la38}) are
canonical equations corresponding to the Hamiltonian function.  Since $E$  does
not explicitly depend on time $t$, its value (which can be called energy) is
conserved. Actually the energy is zero on the instantonic solution since at
$t\to-\infty$ we have $m_{12}\to0$ and $R_{12}\to\infty$. Note that since
the Hamiltonian (\ref{la39}) explicitly depends on the coordinates via $\beta$,
there is no momentum conservation law.

Before proceeding to the solution of the instanton equations, let us make a
remark concerning fluctuations on the background of the instanton. In the
linear approximation over the fluctuations we obtain an estimate for the
typical fluctuation of $R^\gamma$
\begin{equation}
\left(\delta R^\gamma\right)^2
\sim \gamma DR^\gamma |t|{d}^{-1} \,.
\label{fl2} \end{equation}
Note that the fluctuations of $R$ tend to zero when $t\to0$ since $R_{12}$ is 
fixed at $t=0$. Comparing the estimate (\ref{fl2}) with Eq. (\ref{aa40}) we
obtain
\begin{equation}
{\left(\delta R^\gamma\right)^2}/
{R^{2\gamma}}\sim{d}^{-1} \,.
\label{fl3} \end{equation}
We conclude that the fluctuations on the background of our instanton are weak
provided ${d}\gg1$. The above evaluations are rough and need a more
accurate analysis (see Appendix \ref{fluctu}). Nevertheless, they show that
the Richardson behavior (\ref{aa40}) inherent for our instanton suppresses
fluctuations on its background.

The system (\ref{la37}) and (\ref{la38}) consists of two nonlinear
integro-differential equations with boundary conditions imposed on the opposite
sides of the time interval, that is, at $t=0$ for $R_{12}$ and at $t=-\infty$
for $m_{12}$. Therefore, in the general case it is very difficult to solve the
instanton equations. Nevertheless, one can hope that for some particular
objects the system of equations can be reduced to a simpler form allowing the
complete solution. This hope comes true for the structure functions.

\section{Instanton for Structure Functions}
\label{instant}

Using the general scheme developed in Sec. \ref{lagrange} we will examine the
expressions for the structure functions (\ref{str}) at large $n$. In other
words, we will be interested in the statistics of the passive scalar
difference taken at the points separated by the distance ${\bbox r}$. Since the
diffusivity is neglected, we cannot examine the difference
$\theta({\bbox r}/2)-\theta(-{\bbox r}/2)$ itself. Nevertheless, we can
treat the statistics of the differences averaged over separations near
${\bbox r}$. So we should consider the object (\ref{bb14}) with 
\begin{equation}
\beta({\bbox r}_1)=
\delta_\Lambda\left({\bbox r}_1-\frac{{\bbox r}}{2}\right)
-\delta_\Lambda\left({\bbox r}_1+\frac{{\bbox r}}{2}\right) \,.
\label{bl23} \end{equation}
Here $\delta_\Lambda({\bbox r})$ is the function with the width
$\Lambda^{-1}\gg r_d$ satisfying the condition $\int {\rm d}{\bbox
r}\,\delta_\Lambda({\bbox r})=1$, which can be called a smeared
$\delta$ function. Then we can write
\begin{equation}
S_n\approx\langle|\vartheta|^n\rangle
\sim \left.\exp\left(i{\cal I}-n
+n\ln|\vartheta|\right)\right|_{\rm inst} \,,
\label{basic} \end{equation}
where we used Eq. (\ref{as20}) and substituted Eq. (\ref{qq2}).

\subsection{Reduction}
\label{reduc}

Now we turn to the instantonic equations (\ref{la37}) and (\ref{la38}). 
Let us observe, that since the source on the right-hand side of Eq.
(\ref{la38}) is proportional to $\beta({\bbox r}_{1})\beta({\bbox r}_{2})$,
the field $m_{12}$ can be approximated as
\begin{eqnarray} &&
m_{12}=\frac{i m_+}{2}\left\{
\delta_\Lambda\left({\bbox r}_1-\frac{{\bbox r}}{2}\right)
\delta_\Lambda\left({\bbox r}_2-\frac{{\bbox r}}{2}\right)\right.
\nonumber \\ &&
+\left.\delta_\Lambda\left({\bbox r}_1+\frac{{\bbox r}}{2}\right)
\delta_\Lambda\left({\bbox r}_2+\frac{{\bbox r}}{2}\right)\right\} 
\nonumber \\ &&
-\frac{im_-}{2}\left\{
\delta_\Lambda\left({\bbox r}_1-\frac{{\bbox r}}{2}\right)
\delta_\Lambda\left({\bbox r}_2+\frac{{\bbox r}}{2}\right)\right.
\nonumber \\ &&
+\left.\delta_\Lambda\left({\bbox r}_1+\frac{{\bbox r}}{2}\right)
\delta_\Lambda\left({\bbox r}_2-\frac{{\bbox r}}{2}\right)\right\} \,,
\label{la44} \end{eqnarray}
where $m_\pm$ are functions of time only. Writing it, we implicitly assumed
that the field $R_{12}$ is smooth near the points $\pm{\bbox r}/2$.
Then the relations (\ref{aa38}) and (\ref{qq2}) give 
\begin{eqnarray}
\vartheta^2=2n\int\limits_{-\infty}^0 {\rm d}t\,
\left\{\chi\left( R_+\right)-
\chi\left( R_-\right)\right\} \,,
\label{qq3} \end{eqnarray}
where we introduced
\begin{eqnarray} &&
R_+(t)=R(t,{\bbox r}/2,{\bbox r}/2)\,,\
R_-(t)=R(t,{\bbox r}/2,-{\bbox r}/2)\,.
\label{la45} \end{eqnarray}

Substituting the expression (\ref{la44}) into Eqs. (\ref{la37}) and (\ref{la38})
we obtain a closed system of ordinary differential equations for $m_\pm$
and $R_\pm$. It is convenient to proceed in terms of the effective action.
Substituting Eq. (\ref{la44}) into Eq. (\ref{aa60}) we get
\begin{eqnarray} &&
i{\cal I}=\int_{-\infty}^0 \!\!dt\left[\gamma^{-1}\left(
m_-\partial_tR_-^\gamma-m_+\partial_tR_+^\gamma\right)-E\right] \,,
\label{lagr} \\ &&
E=y^2\left\{\chi\left(R_+\right)-
\chi\left(R_-\right)\right\}+D(m_+-m_-) 
\nonumber \\ &&
-\frac{D(2-\gamma)}{4{d}({d}-1)}\left\{
m_-^2\varphi_1+2 m_- m_+\varphi_2+m_+^2\varphi_3 \right\} \,.
\label{qq1} \end{eqnarray}
Here we introduced the designations
\begin{eqnarray} &&
\varphi_1=\frac{4({d}+1-\gamma)}{2-\gamma}R_-^{2\gamma-4}
\left[R_-^{2-\gamma}-R_+^{2-\gamma}
\right]\left[R_-^{2}-R_+^{2}\right]
\nonumber \\ &&
-R_-^{2\gamma-4}\left[R_+^{4-\gamma}
+\frac{(2R_-^{2}-R_+^{2})^2}{R_-^\gamma}\right] \,,
\label{varphi} \\ &&
\varphi_2=R_+^\gamma\left[
\frac{R_+^{2-\gamma}}{R_-^{2-\gamma}}+
\frac{2R_-^2-R_+^2}{R_-^2}\right], \
\varphi_3=-R_+^\gamma\left[
1+\frac{R_+^\gamma}{R_-^\gamma}\right] \,.
\nonumber \end{eqnarray}
Since the effective action (\ref{lagr}) depends only on the functions
$m_\pm(t)$ and $R_\pm(t)$, one can obtain the system of ordinary differential
equations for the functions as extremum conditions of the action. The boundary
conditions for the equations are $R_+=0$ and $R_-=r$ at $t=0$ [see Eq.
(\ref{term})] and $m_\pm\to0$ at $t\to-\infty$.  Resolution of the system
allows one to find $m_\pm$ and $R_\pm$ as functions of time. Once they are
known, it is possible to restore the function $R_{12}$ in the whole space from
Eq. (\ref{la37}). The problem is discussed in Appendix \ref{triangle}. There we
argue that the function $R_{12}$ is really smooth in space which is a
justification of the procedure described.

Since we accept Eq. (\ref{ineq}), ${d}\gg1$. Using the inequality one can
keep in the functions (\ref{varphi}) only the terms of the main order over
${d}$. This means that one can neglect in Eq. (\ref{varphi}) the second
contribution to $\varphi_1$ in comparison to the first one and also
$\varphi_2$, $\varphi_3$ in comparison to $\varphi_1$. Potentially this 
procedure is dangerous. We will show that due to the smallness of $r/L$, the
intervals where $R_--R_+\ll R_-$ play an important role. Then, we see that
it is the difference of $R_\pm$ that enters the first term in $\varphi_1$,
while the others do not contain this smallness. Therefore, we observe
cancellations that could lead to a competition of ${d}$ and $L/r$ (the
latter parameter is considered as the largest in the problem). To check the
possibility, we performed calculations keeping all the terms in Eq.
(\ref{varphi}).
The calculations are sketched in Appendix \ref{corr}. They show that in the
final expressions only combinations of $\varphi_{1,2,3}$, containing the same
cancellations are of importance. The legitimacy of the procedure is proved.  

Omitting $\varphi_{2,3}$ in the expression (\ref{qq1}) and then varying 
the action (\ref{lagr})
over $m_+$, we get a trivial equation for $R_+$,
\begin{equation}
\gamma^{-1}\partial_t R_+^\gamma=-D \,.
\label{nn1} \end{equation}
Its solution, satisfying the boundary condition $R_+(0)=0$ is simply
\begin{equation}
R_+^\gamma=\gamma D|t| \,.
\label{nn2} \end{equation}
To examine the behavior of $R_-$ it is convenient to pass to the new variables
\begin{equation}
R_+=Le^\xi\,, \quad
R_-^\gamma=R_+^\gamma(1+v)\,, \quad
\mu=m_-R_+^\gamma \,.
\label{nn3} \end{equation}
As time $t$ goes from $0$ to $-\infty$, the variable $\xi$ runs from $-\infty$
to $+\infty$ and $v$ runs from $+\infty$ to $0$. The latter is clear from the
asymptotic behavior $R_-^\gamma\approx R_+^\gamma=\gamma D|t|$ at
$t\to-\infty$.  The relation (\ref{qq3}) in terms of the new variables is
\begin{equation}
\vartheta^2=2n\frac{L^\gamma}{D}
\int\limits_{-\infty}^{+\infty} {\rm d}\xi\,
e^{\gamma\xi}\left[
\chi\left(R_+\right)-\chi\left(R_-\right)\right] \,.
\label{nnk} \end{equation}

Recall that the energy $E$ entering the action (\ref{lagr}) is an integral of
motion whose value is equal to zero. Thus we can perform the standard
procedure of excluding a degree of freedom in a canonical system.  Equating the
expression (\ref{qq1}) to zero, we can express $m_+$ in terms of $\mu$, $v$, and
$\xi$. Substituting the result into Eq. (\ref{lagr}), we get
\begin{eqnarray} &&
-i{\cal I}=\int\limits_{-\infty}^{+\infty} {\rm d}\xi\,
(\gamma^{-1}\mu\partial_\xi v-H) \,,
\label{nn4} \\ &&
H\!=\!-\mu v\!+\!\frac{\mu^2}{{d}}\phi(v)
\!+\!\frac{|y|^2 L^\gamma}{D}\left[
\chi\left(R_+\right)\!-\!\chi\left(R_-\right)
\right] e^{\gamma\xi} \,,
\label{ham}  \\ &&
\phi\!=\!(1\!+\!v)^{2-4/\gamma}
\left[(1\!+\!v)^{2/\gamma-1}\!-\!1
\right]\left[(1\!+\!v)^{2/\gamma}\!-\!1\right] \,.
\label{phi} 
\end{eqnarray} 
Here we kept main contributions over ${d}$ only. In Eq. (\ref{ham}) we set
$y^2=-|y|^2$ since as follows from Eqs. (\ref{qq2}) and (\ref{qq3}) $y$ is 
an imaginary number. Extremum conditions for the action (\ref{nn4}) read
\begin{equation}
\gamma^{-1}\frac{{\rm d}v}{{\rm d}\xi}=\frac{\partial H}{\partial\mu} \,, \qquad
\gamma^{-1}\frac{{\rm d}\mu}{{\rm d}\xi}=-\frac{\partial H}{\partial v} \,,
\label{nn7} \end{equation}
which are canonical equations for the variables $\mu$ and $v$ in the
time $\xi$. Of course, Eqs. (\ref{nn7}) could be obtained
directly from the extrema conditions for the action (\ref{lagr}).

To conclude, we reformulated the problem as follows: find such a value of $y$
that the solution of Eqs. (\ref{nn7}) with the given $y$, being substituted
into Eq. (\ref{nnk}), reproduces the correct value of $\vartheta=-in/y$. Below
we discuss the first and the most difficult part of the program that is
solution of the system (\ref{nn7}). Though it cannot be integrated exactly, we
can solve the system approximately by asymptotic matching, which is enough to
determine the structure functions $S_n$.

\subsection{General structure of the instanton}
\label{gener}

The evolution of $R_-$ in 'time' $\xi$ can be divided into three stages. During
the first stage, starting at $\xi=-\infty$, both $R_+$ and $R_-$ are much less
than $L$ and it is possible to substitute both $\chi(R_+)$ and $\chi(R_-)$ by
$\chi(0)$. Then the last term in Eq. (\ref{ham}) is equal to zero. During the
second stage $R_\pm\sim L$ and the last term in Eq. (\ref{ham}) is of
importance. During the final stage, where $R_+\approx R_-\gg L$, one can
again neglect the last term in Eq. (\ref{ham}). Note that only the second stage
contributes to $\vartheta^2$, which can be seen from Eq. (\ref{nnk}). Since the
Hamiltonian $H$ [Eq. (\ref{ham})] does not explicitly depend on `time' $\xi$ during
the first and the third stages, its value is conserved there. Actually,
the value of $H$ is equal to zero during the third stage since $\mu\to0$ and
$R_+\approx R_-\gg L$ at $\xi\to+\infty$. On the other hand, during the first
stage the value $H_1$ of the Hamiltonian function $H$ is nonzero. Therefore,
during the second stage the value of $H$ diminishes and should finally reach
zero when the trivial third stage starts. The value of $H_1$ as a function of
$n$ has to be established from the matching of the stages. 

Now we are going to solve Eq. (\ref{nn7}) for the first stage. 
Resolving the equation $H=H_1$ in terms of $\mu$ we get
\begin{eqnarray} &&
\mu=\frac{{d}}{2\phi}(v-G) \,,
\label{nn8} \\ && 
G(v)=\pm\sqrt{v^2+\frac{4H_1\phi}{{d}}} \,.
\label{nn9} \end{eqnarray}
Then we find from Eq. (\ref{nn7})
\begin{equation}
\gamma^{-1}\frac{{\rm d}v}{{\rm d}\xi}=-G(v) \,.
\label{vvv} \end{equation}
At $\xi\to-\infty$ (that is at small $|t|$) the function $v$ should decrease
with increasing $\xi$ since $R_-\approx r$ and $R_+$ increases. To ensure the
negative value of ${\rm d}v/{\rm d}\xi$ in Eq. (\ref{vvv}) one should take the positive
sign of the square root in Eq. (\ref{nn9}), which leads to a positive value of
$G$. The sign of $G$ can be changed if during the evolution $G$ turns into
zero which corresponds to the presence of a reverse point in the dependence $v$
on $\xi$.  

Equation (\ref{vvv}) enables one to find $v$ as a function of $\xi$. Let us 
integrate the equation over $\xi$ from $-\infty$ to some value. Then we get
\begin{eqnarray} &&
\int^v_\infty {\rm d}x\left[\frac{1}{x}-\frac{1}{G(x)}\right]=
\ln\left[\frac{v R_+^\gamma}{r^{\gamma}}\right] \,.
\label{fv0} \end{eqnarray}
To avoid difficulties related to infinite values of $\xi$ and $v$ at the
initial point, we subtracted from $G^{-1}$ its asymptote $G^{-1}(x)\approx 1/x$
at large $x$.  This enforces the convergence of the integral (\ref{fv0}) at
large $x$. The constant of integration in Eq. (\ref{fv0}) was established from
the limit $v\to\infty$: since the integral on the left-hand side of Eq.
(\ref{fv0}) tends to zero as $v$ increases, the right-hand side of Eq.  (\ref{fv0})
should also tend to zero. This requirement is ensured by the $r$-dependent
factor in Eq. (\ref{fv0}) since $R_+^\gamma\approx r^\gamma/v$ at $v\to\infty$, as
follows from the boundary condition $R_-(0)=r$ and Eq. (\ref{nn3}). The
left-hand side of Eq. (\ref{fv0}) should be viewed as a contour integral,
which determines its value in the case of the non-monotonic behavior of $v$
as a function of $\xi$.

Equation (\ref{fv0}) allows us to establish a relation for the parameters
characterizing the first stage. Let us consider the integral over the whole
first stage. Then we should substitute $v=v_*$ in Eq. (\ref{fv0}), where $v_*$
is the value of $v$ at the end of the first stage. The initial substage (where
$v\gtrsim 1$) gives a constant of order unity in the integral on the
left-hand side of Eq. (\ref{fv0}) since $G(x)\approx x$ there. We neglect the
contribution substituting $v\sim 1$ as the lower limit in the integral. Then
the integral $\int {\rm d}x/x$ produces just $\ln v$, which is canceled by the
corresponding term on the right-hand side. Next, at the boundary between the
first and the second stages $R_-\sim L$ since the pumping enters the game
there. Therefore, with the logarithmic accuracy one can write
\begin{eqnarray} &&
-\int^{v_*}_1\,\frac{{\rm d}x}{G(x)}=
\gamma\ln\left(\frac{L}{r}\right) \,.
\label{log}\end{eqnarray}
We see that there is a large parameter $L/r$ in the argument of the logarithm
on the right-hand side of Eq. (\ref{log}).  An analysis shows that due to this
large parameter there are only two possibilities to satisfy the relation
(\ref{log}). Both of them are related to zeros of the function $G$ because
only near the points where $G$ is small can the integral reach a large value.
The first possibility is realized when $G$ is zero only at $v=0$. In
this case $v_*\ll1$ and $v$ is a monotonically decreasing function. The second
possibility is that $G$ has a zero at some point $v=v_r$.  That is just the
reverse point where the derivative ${\rm d}v/{\rm d}\xi$ changes its sign; see Eq. 
(\ref{vvv}). Then, the integral on the left-hand side of Eq. (\ref{log}) is
determined by the vicinity of the point since $G$ is small there.  

A choice between the possibilities depend on the value of $H_1$.
If $H_1>H_c$, where
\begin{eqnarray} &&
H_c=-\frac{{d}\gamma^2}{8(2-\gamma)} \,,
\label{kn1} \end{eqnarray} 
then $G$ cannot be zero (except for the point $v=0$). Then
the integral on the left-hand side of Eq. (\ref{log}) reaches its large value
at $v\ll1$. Substituting into Eq. (\ref{nn9}) the asymptotic expression
\begin{eqnarray}&&
\phi\approx\frac{2(2-\gamma)}{\gamma^2}v^2 \,,
\label{mk1}\end{eqnarray}
valid at $v\ll1$, we can calculate the integral on the left-hand 
side of Eq. (\ref{log}) with the logarithmic accuracy and find
\begin{eqnarray}
\ln v_*={\gamma\sqrt{1-H_1/H_c}}\,
\ln\left(\frac{r}{L}\right) \,.
\label{aaf}\end{eqnarray}
We see that due to $r\ll L$, indeed $v_*\ll1$.

In the opposite case $H_1<H_c$ the situation is more complicated. From
the asymptotic expression 
\begin{equation}
G^2(v)\approx\left[
-\frac{8(2-\gamma)}{{d}\gamma^2}(H_c-H_1)+
\frac{4-\gamma}{2\gamma}v\right]v^2 \,,
\label{kn2}\end{equation}
valid at $v\ll1$, we see that $G$ is zero at $v=v_r$, where
\begin{equation}
v_r=\frac{16(2-\gamma)}{{d}\gamma(4-\gamma)}(H_c-H_1) \,.
\label{kn3}\end{equation}
It is just the reverse point where the derivative ${\rm d}v/{\rm d}\xi$ changes its sign. 
Therefore, the sign of $G$ is positive during the initial part of the first
period and negative during the final one. Thus we should take the upper sign
in Eq. (\ref{nn9}) for the first part and the lower sign for the second part. The
main contribution to the left-hand side of Eq. (\ref{log}) is determined by the
region near the reverse point $v-v_r\sim v_r$ where we can use the expression
(\ref{kn2}). The explicit integration  gives
\begin{equation}
\sqrt{\frac{{d}}{2(2-\gamma)}}\frac{\pi}{\sqrt{H_c-H_1}} 
=\gamma\ln\frac{L}{r} \,.
\label{kn5}\end{equation}
Since the logarithm is large, $H_1$ is close to $H_c$ and hence $v_r\ll1$,
as we implicitly assumed in the expression (\ref{kn2}). Note that Eq.
(\ref{kn5}) does not fix the value of $v_*$, as it was for $H_1>H_c$. 

Now, we should extract additional relations that, along with Eq. (\ref{aaf}) or
Eq. (\ref{kn5}) will fix the instantonic solution and determine the final
answer for the structure functions. It can be done by establishing the
evolution during the second stage and by its subsequent matching with the first
stage. Unfortunately, the procedure is rather lengthy and is individual for
each particular case. We present the calculations in 
Appendix \ref{solv}.

\subsection{Expressions for structure functions}
\label{results}

Based on the reasoning given in the preceding subsection and on the
calculations described in Appendix \ref{solv}, one can establish expressions
for the structure functions from the relation (\ref{basic}). Here we enumerate
basic results, referring the reader interested in technical details to
Appendix \ref{solv}.

The case $H_1>H_c$ is realized if $n<n_c$ (see Appendix \ref{intermt}), where
\begin{equation}
n_c=\frac{{d}\gamma}{2(2-\gamma)} \,.
\label{crit} \end{equation}
Calculating the action ${\cal I}$ and $\vartheta$ (see Appendix 
\ref{intermt}) and substituting the result into Eq. (\ref{basic}), we obtain
\begin{eqnarray} &&
S_n\sim\left(\frac{n}{\gamma}\frac{P_2 C_1}{D}
L^\gamma\right)^{n/2}\left(\frac{r}{L}\right)^{\zeta_n} \,.
\label{mk11} \\ &&
\zeta_n=\frac{n\gamma}{2}-\frac{(2-\gamma)n^2}{2{d}} \,.
\label{mk12} \end{eqnarray}
The quantity $C_1$ in the expression (\ref{mk11}) is a constant of
order unity, whose value
depends on the shape of $\chi$ (that is on the details of the pumping) and is
consequently nonuniversal. Note that the $r$-independent factor in Eq.
(\ref{mk11}) is determined by the single-point root-mean-square value of the
passive scalar
\begin{equation}
\theta_{\rm rms}^2\sim \frac{P_2}{D\gamma} L^\gamma \,. 
\label{rms} \end{equation}
Comparing the expression (\ref{mk12}) with Eq. (\ref{pert}), we see that they
coincide under the conditions $n\gg1$ and ${d}\gg1$ that were implied
in our derivation. Surprisingly, the $n$ dependence of $\zeta_n$ given by
Eq. (\ref{pert}) is correct not only in the limit (\ref{pert1})
(that is for $n\ll n_c$), but up to $n=n_c$, which is the boundary value for
Eqs. (\ref{mk11}) and (\ref{mk12}).

A detailed consideration of the case $H_1>H_c$ is presented in Appendix
\ref{remote}. 
It shows that this possibility is realized at $n>n_c$. Then, the scaling
exponents $\zeta_n$ appear to be $n$ independent and equal to the value
\begin{equation}
\zeta_c=\frac{{d}\gamma^2}{8(2-\gamma)}.
\label{zc} \end{equation}
The $n$-dependent numerical factors in $S_n$ can be found in two limits:
$n-n_c\ll n_c$ and $n\gg n_c$. The former case is discussed below,
while in the latter case one can obtain (see Appendix \ref{remote})
\begin{eqnarray} &&
S_n\sim\left(\frac{n}{\gamma}\frac{P_2 C_2}{D}
L^\gamma\right)^{n/2}\left(\frac{r}{L}\right)^{\zeta_c} \,.
\label{kkn11} \end{eqnarray}
The quantity $C_2$ in Eq. (\ref{kkn11}) is again a non-universal constant of 
order unity. The expression (\ref{kkn11}) corresponds to the factorized 
Gaussian PDF
\begin{equation}
{\cal P}(\vartheta)\sim \left(\frac{r}{L}\right)^{\zeta_c}
\exp\left(-\frac{\gamma D\vartheta^2}{2C_2P_2L^\gamma}\right)\,. 
\label{remo} \end{equation}
Let us stress that when calculating $S_n\approx\langle|\vartheta|^n\rangle$
with the help of the PDF (\ref{remo}), the characteristic $\vartheta$ is of
the order of the single-point root-mean-square value of the passive scalar (\ref{rms})
and the relatively small value of the result (\ref{kkn11}) compared to a
single-point value is ensured only by the small $r$-dependent factor in Eq.
(\ref{remo}). In Appendix \ref{special} we establish the inequality 
\begin{equation}
\ln\frac{n}{d}<\gamma\ln\frac{L}{r} \,,
\label{jkn12} \end{equation}
which restricts the region where expression (\ref{kkn11}) is correct. For
larger $n$ the character of the PDF essentially changes and it tends to
a single-point PDF that is similar to Eq. (\ref{remo}) but does not contain the
$r$-dependent factor.

Note that the cases $\gamma\ll1$ and $2-\gamma\ll1$ need a special
analysis which is performed in Appendix \ref{special}. The answer (\ref{kkn11})
should be slightly corrected in the case $\gamma\ll1$ and keeps its form at
$2-\gamma\ll 1$.  

We can treat the structure function $S_n$ as a continuous function of $n$. 
Then the vicinity of the critical value $n=n_c$ requires a separate
consideration which is presented in Appendix \ref{critic}. The main peculiarity
that appears in the expressions for the structure functions is a critical
dependence on $n$. The expression for the structure functions can be written as
\begin{equation}
S_n\sim\left[\frac{(n-n_c)^2}{\gamma n_c}\frac{P_2 C_\pm}{D}
L^\gamma\right]^{n_c/2}\left(\frac{r}{L}\right)^{\zeta_n} \,,
\label{strcr} \end{equation}
which implies the condition $|n-n_c|\ll n_c$. The factors $C_\pm$ are
non-universal constants of order unity which are different for the cases
$n<n_c$ and $n>n_c$. The exponents $\zeta_n$ in expression (\ref{strcr}) are
determined by Eq. (\ref{mk12}) if $n<n_c$ and $\zeta_n=\zeta_c$ [Eq. (\ref{zc})] if
$n>n_c$. In the consideration made above we suggested that $r/L$ is the
smallest parameter of our theory. However, if $n\to n_c$, then $|n_c-n|$ starts to
compete with $r/L$ and at small enough $n_c-n$ the consideration presented in
Appendix \ref{critic} is inapplicable.  The criterion that determines the
validity of Eq. (\ref{strcr}) is established in Appendix \ref{critic}
\begin{equation}
\gamma\ln\frac{L}{r}\gg\frac{n_c}{|n-n_c|} \,.
\label{crit1} \end{equation}
We see that the first factor in Eq. (\ref{strcr}) possesses the critical
behavior proportional $|n-n_c|^{n_c}$ that is saturated in the narrow vicinity near
$n=n_c$, where {the} condition (\ref{crit1}) is violated. To avoid a
misunderstanding, let us stress that despite the critical behavior, $S_n$
remains a monotonically increasing function of $n$ at a fixed $L/r$. This is
obvious for $n>n_c$, whereas for $n<n_c$ it accounts for the stronger
dependence on $n$ of the second ($r$-dependent) factor in Eq. (\ref{strcr}), which
is guaranteed by the inequality (\ref{crit1}).

We presented the results of the analysis based on the saddle-point
approximation. The account of fluctuations on the background of our instanton
could, in principle, change the results. Particularly the value of $\zeta_n$
could increase. Therefore, one should estimate the role of the fluctuations. 
The corresponding analysis is presented in Appendix \ref{fluctu}. It shows that
for the condition (\ref{ineq}) fluctuation effects are weak and cannot
essentially change the results obtained.

\section*{Conclusion}

We have performed an investigation of the structure functions in the Kraichnan
model in the framework of the instantonic formalism. Though our approach is
correct only for large dimensionalities of space, we observe a nontrivial
picture, some peculiarities of which could be realized in a wider context.
Below we discuss the results obtained.

We have established the $n$ dependence of the scaling exponents, which are
determined by the expression (\ref{mk12}) for $n<n_c$ and remain the constant
(\ref{zc}) for $n>n_c$ where $n_c$ is defined by Eq. (\ref{crit}). Our results
contradict the schemes proposed in \cite{closure,97Yak}. The value
(\ref{zc}) is different from and smaller than the constant obtained in
\cite{97Chert}, which can be considered really as an estimate from above. For
$n\ll n_c$ our expression coincides with the answer obtained perturbatively
\cite{95CFKLb,95CF} at large ${d}$. Surprisingly, the quadratic dependence
of $\zeta_n$ on $n$ is kept up to $n=n_c$. Such an $n$ dependence of $\zeta_n$ is
well known from the so-called log-normal distribution proposed by Kolmogorov
\cite{62Kol}.  

The expressions (\ref{mk11}) and (\ref{kkn11}) reveal the combinatoric prefactors in
$S_n$ that are characteristic rather of a Gaussian distribution. A natural
explanation can be found in terms of zero mode ideology
\cite{95SS,95GK,95CFKLb,95CF,96BGK,98BGK}. We know that for $n>2$ the main
contribution to the structure function $S_n$ in the convective interval is
related to zero modes of the equation for the $n$th-order correlation function
of the passive scalar. The exponents of the modes are determined by the
equation (and could be very sensitive to the value of $n$), whereas numerical
coefficients before the modes (determining their contribution to $S_n$) have to
be extracted from matching on the pumping scale where the statistics of the
passive scalar is nearly Gaussian. This explains the combinatoric prefactors in
Eqs. (\ref{mk11}) and (\ref{kkn11}). Probably the most striking feature of our results is
the unusual behavior of $S_n$ (treated as continuous functions of $n$) near
$n=n_c$, which is determined by the expression (\ref{strcr}).

Now we briefly discuss the interpretation of our results. The log-normal
answer (\ref{mk11}) and (\ref{mk12}) can be obtained if we accept that for large 
fluctuations, giving the main contribution into the structure function $S_n$,
the pumping is inessential and the fluctuation is smooth on the scale $r$.
Then one obtains from Eq. (\ref{adv}) the equation for the passive
scalar difference taken at the separation ${\bbox r}$,
\begin{equation}
\partial_t\ln(\Delta\theta)=-{\bbox v}{\bbox r}/r^2 \,, 
\nonumber\end{equation}
where we substituted $\nabla\theta$ by $(\Delta\theta)/r$. We immediately get
from this equation a log-normal statistics for $\Delta\theta$ that is a
consequence of the central limiting theorem. The saturation at $n>n_c$ can be
explained by the presence of quasidiscontinuous structures in the field
$\theta$ making the main contribution to the high-order correlation functions
of $\theta$. Note also a similar non-analytical behavior of $\zeta_n$ for
Burgers's turbulence \cite{Frish}, which is explained by the presence of shocks in
the velocity field. Although formally our scheme is applicable only in the
limit $d\gamma\gg1$, one can hope that the main features of our results persist
for arbitrary values of the parameters. This hope is supported by
\cite{98FMVa}, where a saturation of $\zeta_n$ was observed in numerical
simulations of the Kraichnan model at ${d}=3$.

\acknowledgements

We are grateful to G. Falkovich for valuable remarks, and to M. Chertkov, R. 
Kraichnan, D. Khmelnitskii, and M. Vergassola for useful discussions. E. B. 
acknowledges support by the Joseph Meyerhoff Foundation and a grant from
the Israel
Science Foundation. V. L. acknowledges support from the Einstein Center,
from the Minerva Center for Nonlinear Physics at the Weizmann Institute, and
from the ENS-Landau Institute Twinning Programme.

\end{multicols}

\appendix

\begin{multicols}{2}

\section{Single Lagrangian Separation}

In this appendix we treat the statistics of a single Lagrangian separation
defined by Eq. (\ref{la30}). The consideration will allow us to establish
the relation (\ref{aa35}) and also to clarify the condition (\ref{ineq}).

\subsection{Richardson law}
\label{rich}

A single Lagrangian difference ${\bbox R}$ between two Lagrangian
trajectories ${\bbox \varrho}$ and ${\bbox \varrho}+{\bbox R}$
is governed by the equation
\begin{equation}
\partial_t R_\alpha=w_\alpha
\equiv v_\alpha({\bbox\varrho}+{\bbox R})
-v_\alpha({\bbox\varrho}) \,,
\label{la8} \end{equation}
with the correlation function
\begin{eqnarray} &&
\langle w_\alpha(t_1)w_\beta(t_2)\rangle
=2{\cal K}_{\alpha\beta}({\bbox R})\delta(t_1-t_2) \,,
\label{la9} \\ &&
{\cal K}_{\alpha\beta}({\bbox R})\!=\!\frac{D}{{d}}R^{-\gamma}
\left\{\frac{2-\gamma}{({d}-1)}
(R^2\delta_{\alpha\beta}-R_\alpha R_\beta)
\!+\!R^{2}\delta_{\alpha\beta}\right\} \,,
\nonumber \end{eqnarray}
following from Eq. (\ref{eq2}). First of all we get from Eqs. 
(\ref{la33},\ref{la34})
\begin{equation}
\langle\zeta(t_1)\zeta(t_2)\rangle
=\frac{2D}{{d}}R^\gamma\delta(t_1-t_2) \,,
\label{la12} \end{equation}
where in accordance with (\ref{la31}) $\zeta=\gamma^{-1}\partial_t R^{\gamma}$.
Then
\begin{equation}
R(t-\Delta t)-R(t)\approx R^{1-\gamma}(t)
\int\limits_t^{t-\Delta t}{\rm d}\tau\,\zeta(\tau) \,,
\label{mo11} \end{equation}
where we believe $\Delta t>0$ to be a small time interval
(remind that we treat evolution backwards in time). Averaging over the
velocity statistics on the interval from $t$ to $t-\Delta t$ we get from Eqs.
(\ref{la12},\ref{mo11})
\begin{equation}
\langle[R(t-\Delta t)-R]^2\rangle
=\frac{2D}{d}R^{2-\gamma}
\label{mo12} \end{equation}
where $R\equiv R(t)$. 

Let us write now
\begin{equation}
R_\alpha(t-\Delta t)=R_\alpha(t)
+\int\limits_t^{t-\Delta t}{\rm d}\tau\,
w_\alpha[\tau,R_\alpha(\tau)] \,,
\label{mo10} \end{equation}
which is the direct consequence of Eq. (\ref{la8}). Then we find from
Eq. (\ref{mo10})
in the approximation needed for us
\begin{eqnarray} &&
R^2(t-\Delta t)-R^2(t)\approx
2R_\alpha\int\limits_t^{t-\Delta t}{\rm d}\tau\,
w_\alpha(\tau,{\bbox R})
\nonumber \\ &&
+\int\!\!\!\!\!\int\limits_{\!\!\!\!\!t}^{t-\Delta t}{\rm d}\tau\,{\rm d}\tau'\,
w_\alpha(\tau,{\bbox R})w_\alpha(\tau',{\bbox R})
\nonumber \\ &&
+R_\alpha\frac{\partial}{\partial R_\beta}
\int\limits_t^{t-\Delta t}{\rm d}\tau\,
w_\alpha(\tau,{\bbox R})
\int\limits_t^\tau {\rm d}\tau'w_\beta(\tau',{\bbox R}) \,,
\nonumber \end{eqnarray}
where again ${\bbox R}\equiv{\bbox R}(t)$ and we used the incompressibility
condition $\partial{\bbox w}/\partial{\bbox R}=0$. Averaging the expression
over the velocity statistics on the interval from $t$ to $t-\Delta t$ we get
\begin{equation}
\langle R^2(t-\Delta t)-R^2\rangle
\approx 2\frac{D}{d}({d}+2-\gamma)R^{2-\gamma}\Delta t \,,
\label{mo13} \end{equation}
where we used the expressions (\ref{la9}) and taken into account
$\partial{\cal K}_{\alpha\beta}({\bbox R})/\partial R_\alpha=0$. Then we
obtain from Eqs. (\ref{mo12},\ref{mo13})
\begin{equation}
\langle R(t-\Delta t)-R\rangle
\approx \frac{D}{d}({d}+1-\gamma)
R^{1-\gamma}\Delta t \,.
\label{mo14} \end{equation}
Next, we get from the definition (\ref{la31})
\begin{equation}
\langle R^\gamma(t-\Delta t)-R^\gamma\rangle=
-\gamma\langle\zeta\rangle\Delta t \,.
\label{mo15} \end{equation}
Expanding here the difference up to the second order over $R(t-\Delta t)-R$
and substituting then (\ref{mo12},\ref{mo14}) we find finally
\begin{equation}
\langle\zeta\rangle=-D \,.
\label{mo55} \end{equation}
Note that the average is negative which is a consequence of considering
an evolution backwards in time.

The average value $\langle\zeta\rangle$ is obviously the same for all the
Lagrangian separations. Therefore, we arrive at Eq. (\ref{aa35}) leading
then to Eq. (\ref{aa40}), which is a manifestation of the Richardson law.

\subsection{Simultaneous PDF}
\label{simul}

The Kraichnan model admits a closed description of the simultaneous probability
distribution function ${\cal P}({\bbox R})$ for any single Lagrangian
difference ${\bbox R}$. The point is that Eq. (\ref{la8}) in this case
can be considered as a stochastic process with white noise on the right-hand
side. It is well known how to obtain the equation for ${\cal P}({\bbox R})$ in
the situation. Using the expression (\ref{la9}) we get
\begin{equation}
\partial{\cal P}/\partial|t| ={\cal K}_{\alpha\beta}({\bbox R})
\nabla_\alpha\nabla_\beta {\cal P} \,.
\label{r1} \end{equation}
The same problem for $\gamma=2/3$ was considered by Kraichnan \cite{66Kra-b}
who obtained the asymptotic behavior of ${\cal P}$ at large times $t$. Here we
present a straightforward generalization of Kraichnan's scheme for arbitrary 
$\gamma$.

Suppose that at $t=0$ the PDF is ${\cal P}=R_0^{1-d}\delta(R-R_0)$, the initial
condition corresponds to a fixed value of $|{\bbox R}|$ at $t=0$.  The above
expression implies the normalization condition $\int{\rm d}R\, R^{d-1}{\cal
P}(t,R)=1$.  Solutions corresponding to other initial conditions can be
expressed via this fundamental solution since the equation (\ref{r2}) is
linear. Due to isotropy, ${\cal P}$ will be a function of $R$ only. Then the
equation (\ref{r1}) is rewritten as
\begin{equation}
\frac{d}{D}\frac{\partial {\cal P}}{\partial |t|}
=\frac{1}{R^{d-1}}\frac{\partial}{\partial R}
\left(R^{d+1-\gamma}\frac{\partial}{\partial R}{\cal P}\right) \,.
\label{r2} \end{equation}
Of course (\ref{r2}) can also be obtained directly from Eqs. 
(\ref{la12},\ref{mo55}).  

Performing Laplace transform of Eq. (\ref{r2}), one obtains
\begin{equation}
{\cal P}(t,R)=\int\limits_{A-i\infty}^{A+i\infty}
\frac{{\rm d}\lambda}{2\pi i}
\exp\left(\frac{D}{d}\lambda |t|\right){\cal S}(\lambda, R)\,,
\label{r3} \end{equation}
where all the singularities of ${\cal S}(\lambda)$ have to be to the left
of the integration contour. The function ${\cal S}$ in (\ref{r3})
satisfies the equation
\begin{eqnarray} &&
\lambda{\cal S}-\frac{1}{R^{d-1}}\frac{\partial}{\partial R}
\left(R^{d+1-\gamma}\frac{\partial}{\partial R}{\cal S}\right)
=\frac{1}{R_0^{d-1}}\delta(R-R_0) \,.
\nonumber \end{eqnarray}
The equation can be solved  separately in the regions $R<R_0$ and $R>R_0$ where
we deal with the homogeneous equation and then the matching conditions at
$R=R_0$ give us coefficients. Assuming suitable boundary conditions (${\cal
S}$ is finite at $R\to0$ and at $R\to\infty$) we get
\begin{eqnarray} &&
{\cal S}(\lambda,R)=\frac{2}{\gamma}
(RR_0)^{-d/2+\gamma/2}
\label{r5} \\ &&
\times\left\{
\begin{array}{c}
{\sl I}_\nu\left(\frac{2\sqrt\lambda}{\gamma}R^{\gamma/2}\right)
{\sl K}_\nu\left(\frac{2\sqrt\lambda}{\gamma}R_0^{\gamma/2}\right) 
\quad {\rm if}\quad R<R_0 \,, \\ 
{\sl K}_\nu\left(\frac{2\sqrt\lambda}{\gamma}R^{\gamma/2}\right)
{\sl I}_\nu\left(\frac{2\sqrt\lambda}{\gamma}R_0^{\gamma/2}\right)
\quad {\rm if}\quad R>R_0 \,.
\end{array} \right.
\nonumber \end{eqnarray}
Here, $\nu=d/\gamma-1$ and ${\sl I}_\nu$ is the modified Bessel function,
${\sl K}_\nu$ is the McDonald function.

If we are interested in the asymptotic behavior of ${\cal P}(t,R)$ at large
times $|t|\gg{d}R_0^\gamma/(D\gamma^2)$ then $R$ near the maximum of ${\cal
P}$ satisfy $R\gg R_0$. In addition, in the integral (\ref{r3}) the characteristic
value $\lambda_{\rm char}\propto|t|^{-1}$ is small. Thus we can take only the
second term in (\ref{r5}) (corresponding to $R>R_0$) and substitute the first
term of the expansion ${\sl I}_\nu(z)\simeq \Gamma^{-1}(1+\nu)(z/2)^\nu$. Then
the integral (\ref{r3}) can be taken explicitly and we obtain 
\begin{eqnarray} 
{\cal P}(t,R) R^{d-1}{\rm d}R &=&
\frac{1}{\Gamma(1+\nu)} {\rm d}\xi\,\xi^\nu \exp(-\xi) \,,
\label{r6} \\ &&
\xi=\frac{{d}R^\gamma}{D\gamma^2 |t|}\,, 
\label{r7} \end{eqnarray}
which gives the asymptotic self-similar behavior of PDF. We see that $\ln{\cal
P}\propto-R^\gamma$. It is interesting that the asymptotic PDF is
Gaussian at $\gamma=2$ for any $d$. As we move $\gamma$ from two to zero, the
PDF is getting more and more non-Gaussian reaching an extreme non-Gaussianity
(log-normality) at $\gamma=0$. Note an obvious consequence of Eqs.
(\ref{r6},\ref{r7}):
\begin{eqnarray}
\langle R^{\mu\gamma}\rangle
\propto |t|^{\mu} \,,
\label{rr7} \end{eqnarray}
for arbitrary $\mu$. The asymptotic law (\ref{rr7}) is a manifestation
of the Richardson law.

It is not very difficult to calculate first moments using the expression
(\ref{r6}):
\begin{eqnarray} &&
\langle R^2\rangle=
\left(\frac{D\gamma^2|t|}{d}\right)^{2/\gamma}
\frac{\Gamma(1+\nu+2/\gamma)}{\Gamma(1+\nu)}\,,
\label{r8} \\ &&
\langle R^4\rangle=
\left(\frac{D\gamma^2|t|}{d}\right)^{4/\gamma}
\frac{\Gamma(1+\nu+4/\gamma)}{\Gamma(1+\nu)}\,.
\label{r9} \end{eqnarray}
In the large ${d}$ limit the expressions (\ref{r8},\ref{r9}) give
\begin{eqnarray} &&
\langle R^2\rangle=(D\gamma |t|)^{2/\gamma} \,,
\label{r10} \\ &&
\frac{\langle R^4\rangle-\langle R^2\rangle^2}
{\langle R^2\rangle^2}=
\exp\left(\frac{4}{\gamma d}\right)-1\,,
\label{r11} \end{eqnarray}
irrespective to the value of $\gamma$. We see that fluctuations of $R^2$ are
small if $\gamma d\gg1$ and in the opposite limit $\gamma d\ll1$ the cumulant
of $R^2$ is much larger than $\langle R^2\rangle^2$. We conclude that the 
inequality (\ref{ineq}) is just the condition at which $R^\gamma$ only weakly
fluctuates near the value (\ref{aa40}).

\section{}

Here we discuss the applicability conditions of our scheme. First, the triangle
inequality (\ref{trian}) should be satisfied for our instantonic solution.
Second, fluctuations on the background of our instanton should be weak.

\subsection{Triangle inequality}
\label{triangle}

Here we discuss the triangle inequality (\ref{trian}) that was ignored in our
saddle-point calculations. We argue that if the parameter $d\gamma$ is large,
the inequality (\ref{trian}) is satisfied for the instanton solution. In other
words, we can say that calculating the path integral (\ref{bb57}), we can
dismiss the restriction supplied by (\ref{trian}), because the contribution
from the regions where the inequality is violated, is small in this limit.  

Let us recall that the inequalities (\ref{trian}) are obviously satisfied both
for the initial condition (\ref{term}) and the asymptotic behavior
(\ref{aa40}). Therefore, they could be violated only for times of the order of
the instanton lifetime. Our project will be as follows.  First, we check that
the inequality (\ref{trian}) holds for the instanton solution if we consider the
instantonic equations in the main order over $1/{d}$. Then, we will show, that
next order terms over $1/{d}$ considered in Appendix \ref{corr} can lead to
violation of the inequality if ${d}\gamma\lesssim 1$. Generally, this
procedure is complicated, and here we present only some calculations which
serve as a basis for our conclusions.

As a first step, we have to restore the field $R_{12}$ in the whole space,
that is for any two points ${\bbox r}_1$ and ${\bbox r}_2$. 
This can be done as follows. Let us consider Eq. (\ref{la37}). Substituting
$m_{12}$ in the form (\ref{la44}) we get
\begin{eqnarray} &&
\gamma^{-1}\partial_t R_{12}^\gamma+D=-\frac{Dm_-}{2{d}} 
\label{tri1} \\ &&
\frac{\left(R_{1-}^{2-\gamma}\!\!+\!\!R_{2+}^{2-\gamma}
\!\!-\!\!R_{1+}^{2-\gamma}\!\!-\!\!R_{2-}^{2-\gamma}\right)
\left(R_{1-}^{2}\!\!+\!\!R_{2+}^{2}\!\!-\!\!R_{1+}^{2}\!\!
-\!\!R_{2-}^{2}\right)}{R_{12}^{2-\gamma}R_-^{2-\gamma}} \,.
\nonumber \end{eqnarray}
We kept only the main term over $1/{d}$ in $Q$ (\ref{la34}).
In Eq. (\ref{tri1}) we introduced auxiliary fields
\begin{eqnarray}&&
R_{1+}=R\left({\bbox r}_1,{\bbox r}/{2}\right)\,,\quad 
R_{1-}=R\left({\bbox r}_1,-{\bbox r}/{2}\right) \,.
\label{tri0} \end{eqnarray}
They satisfy the closed system of equations

\end{multicols}

\begin{eqnarray}&&
\gamma^{-1}\partial_t R_{1+}^\gamma+D=-
\frac{Dm_-}{2{d}} \frac{\left(
R_{1-}^{2-\gamma}+R_{+}^{2-\gamma}-R_{1+}^{2-\gamma}-R_{-}^{2-\gamma}
\right)\left(
R_{1-}^{2}+R_{+}^{2}-R_{1+}^{2}-R_{-}^{2}
\right)}{R_{1+}^{2-\gamma}R_-^{-\gamma}} \,,
\label{tri2} \\ &&
\gamma^{-1}\partial_t R_{1-}^\gamma+D=-
\frac{Dm_-}{2{d}} \frac{\left(
R_{1-}^{2-\gamma}+R_{-}^{2-\gamma}-R_{1+}^{2-\gamma}-R_{+}^{2-\gamma}
\right)\left(R_{1-}^{2}+R_{-}^{2}-R_{1+}^{2}-R_{+}^{2}
\right)}{R_{1-}^{2-\gamma}R_-^{2-\gamma}}\,,
\label{tri3}\end{eqnarray}
\begin{multicols}{2}

\noindent
which can be obtained from Eq. (\ref{tri1}). Thus one can restore the function
$R_{12}$ in two steps: first solving Eqs. (\ref{tri2},\ref{tri3}) and then
substituting the functions $R_{1\pm}$ into Eq. (\ref{tri1}). Though at
each step one should solve ordinary differential equations, this is a hard
program and we are not able to perform it entirely. Nevertheless, we can
examine some crucial cases.  

First of all, using the above scheme one can establish the behavior of the
field $R_{12}$ where ${\bbox r}_{1,2}$ are close to $\pm{\bbox r}/2$. The
analysis shows that the field $R_{12}$ is smooth near the points. This is the
justification of our reduction procedure leading to the effective action
(\ref{lagr}).

Simple consideration shows that the most dangerous geometry that could lead to
violation of the triangle inequalities (\ref{trian}) is realized if two of
the three points are close to $\pm{\bbox r}/2$ whereas the third one (say,
${\bbox r}_1$) lies in the middle between the points. In this case
$R_{1+}=R_{1-}$ and the triangle inequality tells in this case
that the difference $2R_{1-}-R_-$ should be positive. To prove the
inequality let us write the equation for this quantity
\begin{eqnarray} &&
\frac{{\rm d}(2R_{1-}-R_-)}{{\rm d}t}=
-D\left(2R_{1-}^{1-\gamma}-R_-^{1-\gamma}\right)
\label{tri4} \\ &&
-\frac{Dm_-}{{d}}\frac{(R_+^{2-\gamma}-R_-^{2-\gamma})(R_+^2-R_-^2)}
{R_-^{2-\gamma}}\frac{\left(R_--2R_{1-}\right)}{R_-R_{1-}} \,.
\nonumber \end{eqnarray}
It is easy to show that at small times the quantity $2R_{1-}-R_-$ increases.
Suppose that at some moment of time it becomes zero. Then, the second term in
the right-hand side of (\ref{tri4}) is zero, while the first one is negative. 
It means that the difference $2R_{1-}-R_-$ increases (recall that we move
backwards in time), though it should approach zero from the positive side.
The contradiction proves that $2R_{1-}-R_-$ is always positive.

Let us now restore the terms subleading over $1/{d}$. Adding the terms to Eq.
(\ref{tri1}) we can repeat our consideration. Again, we should consider the
same geometry as above. Writing the equation for $2R_{1-}-R_-$, we are
convinced that in addition to the two terms presented in Eq. (\ref{tri4}) one
should take into account also the term of the next order over $1/{d}$. This
term is non-zero when $2R_{1-}=R_-$, therefore it starts to compete with the
leading term $-DR_-^{1-\gamma}(2^\gamma-1)$ if $2R_{1-}-R_-$ is small. In this
case the leading term is proportional to $2^\gamma-1$ that is to $\gamma$ at
small $\gamma$. Therefore, if $\gamma {d}$ is not large, we can not make
definite conclusion about the sign of the difference. Thus we arrive at the
inequality (\ref{ineq}) formulated in the main text.  The crucial cases
investigated above make us confident that the triangle inequality holds for our
instanton in the whole space provided the inequality (\ref{ineq}) is satisfied.

\subsection{Fluctuations}
\label{fluctu}

Here we extend the analysis of the fluctuations presented in the Sec.
\ref{genin} that is suitable for all the points excluding vicinities of
$\pm{\bbox r}/2$. The reason is that the quantity $v$ introduced by Eq.
(\ref{nn3}) is small during some stages of the evolution. Therefore,
$R_+=R({\bbox r}/2,{\bbox r}/2)$ and $R_-=R({\bbox r}/2,-{\bbox r}/2)$ are
almost equal to each other. Thus, we should check that fluctuations do not
destroy this proximity. Having the problem in mind we will assume $v\ll1$
below.
 
We can take the effective action (\ref{aa60}) as the starting point of our 
analysis. It will be enough for our purpose to examine fluctuation effects
in the harmonic approximation. Therefore, we should expand the effective
action (\ref{aa60}) up to the second order over the fluctuations $\delta
R_{12}^\gamma$ and $\delta m_{12}$. The first order term of the expansion 
vanishes due to our saddle-point equations. The second order term can be 
written as
\begin{eqnarray} && 
i{\cal I}^{(2)}=i\int {\rm d}t\, {\rm d}{\bbox r}_1\, {\rm d}{\bbox r}_2\,
\delta m_{12}\gamma^{-1}\partial_t\delta R_{12}^\gamma 
\nonumber \\ && 
-\frac{D}{d}\int {\rm d}t\,{\rm d}^4{\bbox r}\, 
Q_{12,34}\delta m_{12}\delta m_{34} 
\label{fl7} \\ && 
+\frac{1}{2}\int {\rm d}t\, {\rm d}{\bbox r}_1\,{\rm d}{\bbox r}_2\, 
|y|^2\chi^{(2)}(R_{12})\beta({\bbox r}_1)\beta({\bbox r}_2) 
\nonumber \\ && 
-\frac{D}{d}\int {\rm d}t\, {\rm d}^4{\bbox r}\, \left\{ 
2\delta Q_{12,34} m_{12}\delta m_{34} 
+Q^{(2)}_{12,34}m_{12}m_{34} \right\} \,.  
\nonumber \end{eqnarray}
where ${\rm d}^4{\bbox r}={\rm d}{\bbox r}_1\,{\rm d}{\bbox r}_2\, 
{\rm d}{\bbox r}_3\,{\rm d}{\bbox r}_4$.
Here $R_{12}$ is the field corresponding to the instanton (recall, that the way
to restore it in the whole space was discussed in Appendix \ref{triangle}),
$m_{12}$ is determined by the expression (\ref{la44}),
and $\chi^{(2)},Q^{(2)}$ are
the second order terms in the expansion over $\delta R_{12}^\gamma$. As follows
from Eqs. (\ref{bl23},\ref{la44}), the last two terms in (\ref{fl7}) are
relevant only for the points close to $\pm{\bbox r}/2$. Therefore, in the
general case we can disregard these terms, and return to the estimate
(\ref{fl3}), which can be obtained if only the two first terms in the action
(\ref{fl7}) are kept. However, we are interested just in the behavior of
$R_{12}$ when the points ${\bbox r}_1$ and ${\bbox r}_2$ are close to
$\pm{\bbox r}/2$. In this case a special analysis is needed.

First, note that short-scale fluctuations of $R_{12}$ are weak due to the
restriction (\ref{trian}) which makes the amplitude of the
fluctuations  proportional to the scale. In other words, we should deal only
with smooth functions $R_{12}$. Next, due to the presence of the second term in
the effective action (\ref{fl7}) fluctuations of $m_{12}$ are relatively
suppressed for points that are not very close to $\pm{\bbox r}/2$. That is a
consequence of the ${\bbox r}$-dependence of $Q_{12,34}$ [Eq. (\ref{la34})]
which in
the case $v\ll1$  has deep minima if the points ${\bbox r}_i$ are close to
$\pm{\bbox r}/2$ (more precisely some linear combinations have deep minima,
they just determine the structure of strong fluctuations of $m$). Therefore,
relevant fluctuations of $R_{12}$ can be estimated in terms of the expression
(\ref{la44}) and, consequently, in terms of the reduced action (\ref{lagr}).
Since in the main approximation over ${d}$ the terms with $\varphi_2$ and
$\varphi_3$ can be neglected in (\ref{qq1}), the integration over $m_ +$ can be
done explicitly. That leads simply to fixing the expression (\ref{nn2}) and
reduces the action to the form (\ref{nn4}).

So we should estimate fluctuations of $v$ and $\mu$ starting from the action
(\ref{nn4}) with the Hamiltonian (\ref{ham}). Actually, we should check the
validity of the semiclassical approximation for this system with one degree of
freedom. It is more convenient to perform this conventional procedure in terms
of the canonically conjugated variables $p$ and $q$ where $p=\mu v/\gamma$ and
$v=\exp(q)$.  Then the Hamiltonian (\ref{ham}) is rewritten as
\begin{eqnarray}&&
H=-\gamma p+\frac{2(2-\gamma)}{d}\left[
1-\frac{4-\gamma}{2\gamma}\exp(q)\right]p^2\nonumber\\&&
-\frac{|y|^2}{D\gamma}R_+^{1+\gamma}\chi'(R_+)\exp(q) \,,
\label{fl9} \end{eqnarray}
where we took into account $v\ll1$. The subsequent analysis shows that the
semi-classical approximation is broken only in the vicinity of the reverse
point $v_r$ that exists at $n>n_c$. Let us estimate this vicinity. Near
the reverse point the last term in (\ref{fl9}) can be neglected and 
resolving the relation $H=H_1$ we get
$$ p=\frac{n_c}{2}\left[1\pm\sqrt\frac{4-\gamma}{2\gamma}
\sqrt{\exp(q)-v_r}\right] \,, $$
where we substituted (\ref{kn3}). The semi-classical approximation is
broken if $p^{-2}dp/dq\sim 1$ which gives
$$ v-v_r\sim \frac{v_r^2}{n_c^2}\ll v_r  \,. $$
Since the main contribution to the integrals such as Eq. (\ref{log}) is made
at $v-v_r\sim v_r$, the above narrow vicinity (where the semi-classical
approximation is broken) is irrelevant for our results.

Of course, the above analysis of the fluctuations on the background of
our instanton is not exhaustive. Nevertheless, we believe that the 
arguments presented demonstrate the weakness of the fluctuations.

\section{}
\label{corr}

As was mentioned in Sec. \ref{reduc}, keeping only the main over $1/{d}$
term in the action (\ref{lagr}) is potentially dangerous because of a possible
conflict of the limits ${d}\gg 1$ and $L/r\gg 1$. If this were the case,
our results
would not be applicable deep in the convective interval, where the ratio $r/L$
is very small. Fortunately, this is not the case. In this appendix we will
consider the equations, following from the action (\ref{lagr}), not neglecting
terms subleading over $1/{d}$. We show that $r/L$ does not interfere with
$1/{d}$, and the scheme, presented in Sec. \ref{instant} is reproduced with
minor modifications.  Repeating the scheme, we obtain the expressions for the
anomalous exponents $\zeta_n$ in this formulation. To avoid a misunderstanding
let us stress that our scheme is correct only in the limit ${d}\gamma>1$
(see Appendix \ref{triangle}). Therefore, one could consider this appendix only
as a demonstration of the absence of the aforementioned conflict of limits.

Below, we will assume that $D=1$, $L=1$, and $P_2=1$ which can be done by
rescaling time $t$, the coordinates ${\bbox r}$, and the passive scalar 
$\theta$. To restore the full answers one should add simply the factor
$P_2/DL^\gamma$ to $\vartheta^2$ in all expressions.

Extremum conditions for the action (\ref{lagr}) give four equations of motion
for the quantities $R_\pm$ and $m_\pm$. As before, the boundary conditions to
the equations are $R_+=0$ and $R_-=r$ at $t=0$, and $m_\pm\to 0$ at
$t\to-\infty$. The equations are canonical, with Hamiltonian given by Eq.
(\ref{qq1}). Since it does not depend on time explicitly, the energy is
conserved. Moreover $E=0$, which is a consequence of the asymptotic behavior
$m_\pm\to 0$ of the instantonic solution at $t\to-\infty$.  

As it is known from classical mechanics, for a Hamiltonian system
which has an integral of motion one can reduce the number of
degrees of freedom by one. In our case, we can pass from the system of
four canonical equations to that of two. Let us express say, $m_+$ via 
the other variables with the help of the conservation law $E=0$:
\begin{eqnarray} &&
m_+{\varphi_3}=\alpha-m_-\varphi_2
\label{mpl} \\ &&
-\sqrt{(m_-\varphi_2-\alpha)^2-
\varphi_3\left[m_-^2\varphi_1+2\alpha(m_-+|y|^2U)\right]} \,,
\nonumber \end{eqnarray}
where
\begin{equation}
U=\chi\left(R_+\right)-\chi\left(R_-\right)\,, \quad
\alpha=\frac{2{d}({d}-1)}{2-\gamma} \,.
\label{en2} \end{equation}
The sign in front of the square root in (\ref{mpl}) should be minus
to ensure the correct behavior of $R_+$ at small time.

Let us make the substitution (\ref{nn3}). Then we obtain a generalization of
Eqs. (\ref{nn4},\ref{ham})
\begin{eqnarray} &&
-i{\cal I}=\int {\rm d}\xi\,\left[\gamma^{-1}\mu\partial_\xi v-H\right]\,,
\label{newac} \\ &&
H=-\mu (1+v)
+\frac{\alpha-\mu\phi_2}{\phi_3}
\label{energ2} \\ &&
-{\phi_3}^{-1}{\sqrt{(\mu\phi_2-\alpha)^2-
\phi_3\left\{\mu^2\phi_1+2\alpha
\left[\mu+|y|^2Ue^{\gamma\xi}\right]\right\}}}
\nonumber \end{eqnarray}
Note that
\begin{eqnarray} &&
U\!=\!\chi(e^\xi)-\chi\left[(1+v)^{1/\gamma}e^\xi\right]\,, \qquad
m_+=\frac{H+\mu(1+v)}{e^{\gamma\xi}}\,.
\nonumber \end{eqnarray}
We introduced here the notations
\begin{eqnarray} &&
\phi_1=-(1+v)^{2-4/\gamma}\left[
1+\frac{(2(1+v)^{2/\gamma}-1)^2}{1+v}\right] 
\label{ss2} \\ &&
+\frac{4({d}+1-\gamma)}{2-\gamma}
\!\left[(1+v)^{1-2/\gamma}\!\!-1\right]
\!\left[(1+v)^{1-2/\gamma}\!\!-1-v\right],
\nonumber \\ &&
\phi_2=2+\frac{1}{(1+v)^{2/\gamma-1}}
-\frac{1}{(1+v)^{2/\gamma}},\quad
\phi_3=-\frac{2+v}{1+v}\,.
\nonumber \end{eqnarray}
One can derive also a generalization of Eq. (\ref{nnk})
\begin{eqnarray} &&
\vartheta^2=\int\limits_{-\infty}^\infty
\frac{2n\alpha \,e^{\gamma\xi} U{\rm d}\xi}
{\sqrt{(\mu\phi_2-\alpha)^2-
\phi_3\left\{\mu^2\phi_1+2\alpha
\left[\mu+|y|^2Ue^{\gamma\xi}\right]\right\}}}
\nonumber
\end{eqnarray}

As before (see Sec. \ref{reduc}) we can divide the evolution into three stages.
Let us analyze the first stage. Since $U=0$ there, the quantity $H$ does not
explicitly depend on $\xi$ and therefore its value (which we designate $H_1$)
is conserved during the first stage. Then from the relation (\ref{energ2}) we
can express $\mu$ via $v$ as follows
\begin{eqnarray} &&
\mu=\frac{\alpha v-H_1S_4\pm F_1}{S_1}
\label{FV} \\ &&
F_1=\sqrt{H_1^2S_3+
2\alpha H_1S_2+\alpha^2v^2}\,.
\label{FV1} \end{eqnarray}
We introduced the shorthand notations
\begin{eqnarray} &&
S_1=\phi_1+2(1+v)\phi_2+\phi_3 (1+v)^2\,,
\label{s1} \\ &&
S_2=\phi_1+\phi_2+(1+v)(\phi_2+\phi_3)\,,
\label{s2} \\ &&
S_3=\phi_2^2-\phi_1\phi_3\,,
\label{s3} \\ &&
S_4=\phi_2+(1+v)\phi_3\,.
\label{s4} \end{eqnarray}
Then we can derive an equation for $v$
\begin{eqnarray} &&
\gamma^{-1}\partial_\xi v=\frac{\pm S_1F_1}
{\alpha S_2+H_1 S_3\mp S_4F_1}\equiv -G(v)
\label{DFR}\end{eqnarray}
This equation is the direct generalization of Eq. (\ref{vvv}).
To ensure the finite value of the action (\ref{newac}),
one should take the lower signs in (\ref{FV},\ref{DFR}). A solution of
Eq. (\ref{DFR}) with the correct boundary condition is
given by Eq. (\ref{fv0}), where $G$ should be substituted from Eq. (\ref{DFR}).

Let us establish the behavior of $G$ at small $v$ which is important
for description of the first stage. To do that, we should take 
into account the relations
\begin{eqnarray} &&
\phi_1=-2-3v+\frac{8d-4\gamma-\gamma^2}{\gamma^2}v^2 \,, 
\nonumber \\ &&
\phi_2=2+v-\frac{2}{\gamma}v^2, \quad
\phi_3=-2+v-v^2 \,,
\nonumber \\ &&
S_1=\frac{8({d}-\gamma)}{\gamma^2}v^2 \,, \quad
S_2=\frac{8({d}-\gamma)}{\gamma^2}v^2 \,, 
\nonumber \\ &&
S_3=\frac{16({d}-\gamma)}{\gamma^2}v^2 \,, \quad
S_4=-\frac{2}{\gamma}v^2 \,,
\nonumber \end{eqnarray}
valid at $v\ll1$. Note that all the functions $S_i$ have the homogeneous
behavior $\propto v^2$ at small $v$. Therefore, starting from relation
(\ref{fv0}) we can find the same estimates (\ref{aaf},\ref{kn3}) for $v_*$ 
and $v_r$  with small corrections of the order $1/{d}$. 

Next, we should analyze the second stage and match it with the first
stage. That is a repetition of the procedure described in Appendix \ref{solv}.
As a result we find the exponents
\begin{eqnarray} &&
\zeta_n=\frac{1}{2}\left(\alpha-
\sqrt{\alpha^2-2\alpha\gamma n+4({d}-\gamma)n^2}\right) \,,
\label{corr1} \\ && 
\zeta_c=\frac{\alpha}{2}
\left(1-\sqrt{1-\frac{\gamma^2}{4({d}-\gamma)}}\right) \,.
\label{corr2} \end{eqnarray}
In the limit $\gamma{d}\gg 1$ we recover the previous results 
(\ref{mk12}) and (\ref{zc}). To avoid a misunderstanding, let us stress that
the expressions (\ref{corr1},\ref{corr2}) cannot be used to establish
$1/{d}$ corrections to the exponents (\ref{mk12},\ref{zc}) since
contributions of the same order, related to the triangle inequalities
(\ref{trian}) and fluctuations, are unknown.

\section{}
\label{solv}

In this appendix we present a consideration of the second stage and matching
conditions for the instanton solution. The procedure appears to depend strongly
on the order $n$ of the structure function. Therefore, we consider
different cases separately . The designations used below were introduced
in Sec. \ref{instant}.

\subsection{Intermediate tail}
\label{intermt}

Let us first consider the case where the behavior of the function $v(\xi)$ is
monotonic during the whole first stage. As was demonstrated in the main body of
the text, this case is realized if $H_1>H_c$ and that the variable $v$ is small
at the end of the first stage. We will show that the variable remains small
also during the second stage, going to zero at the third stage. Therefore, the
evolution of $v$ during the last substage of the first stage and during the
second and the third stages can be described in terms of the Hamiltonian
(\ref{ham}), where the condition $v\ll1$ is utilized, which simplifies the
analysis essentially.

In the limit $v\ll1$ we get from Eqs. (\ref{ham},\ref{phi})
\begin{eqnarray} &&
H=-\mu v+\frac{2(2-\gamma)}{{d}\gamma^2}\mu^2 v^2
-\frac{|y|^2}{\gamma D}R_+^{1+\gamma}\chi'(R_+)v \,.
\label{mk2} \end{eqnarray}
We see that the first two terms in the expression (\ref{mk2}) depend
only on $\mu v$. This is the reason why the equation for the quantity  
\begin{equation}
\frac{{\rm d}(\mu v)}{{\rm d}\xi}=
\frac{|y|^2}{D}R_+^{1+\gamma}\chi'(R_+)v \,, 
\label{mmk3} \end{equation}
that can be derived from Eqs. (\ref{mk2},\ref{nn7}) contains in the right-hand
side only the term proportional to $\chi'(R_+)$. The term can be neglected 
if either $R_+\ll1$ or $R_+\gg1$ that is during the first and during the
third stages. Therefore, the quantity $\mu v$ is conserved there.
Recall that due to the boundary condition $\mu=0$ at $\xi\to\infty$ 
and consequently $\mu v=0$ during the third stage.

Integrating relation (\ref{mmk3}) from any $\xi$ corresponding to
the last substage of the first stage up to $+\infty$ we get the following
integral relation
\begin{equation}
\mu v=-\frac{|y|^2}{D}
\int {\rm d}\xi\, R_+^{1+\gamma}\chi'(R_+)v \,,
\label{mk3} \end{equation}
which is correct for the first stage if $v\ll1$. It is instructive to 
compare Eq. (\ref{mk3}) with the relation
\begin{equation}
\vartheta^2=-\frac{2n}{\gamma D}
\int {\rm d}\xi\, R_+^{1+\gamma}\chi'(R_+)v \,,
\label{mk4} \end{equation}
which can be obtained from Eq. (\ref{nnk}) at $v\ll1$.
Recalling also relation $y=-in/\vartheta$ (\ref{qq2}), we get
\begin{equation}
\mu v=\frac{\gamma n}{2} \,.
\label{mk6} \end{equation}
Substituting Eq. (\ref{mk6}) into (\ref{mk2})
and neglecting the term with $\chi'$
we find that the value of the Hamiltonian during the first stage is
\begin{equation}
H_1=-\frac{\gamma n}{2}
+\frac{2-\gamma}{2{d}}n^2 \,.
\label{mk7} \end{equation}

During the second stage $\mu v$ diminishes from $\gamma n/2$ to zero.
Therefore, the equation 
$$ \frac{{\rm d}\ln v}{{\rm d}\xi}=\gamma
\left(-1+\frac{4(2-\gamma)}{{d}\gamma^2}\mu v\right) \,, $$
shows that the quantity $v$ does not vary essentially during the second 
stage. Then we obtain from Eqs. (\ref{mk4},\ref{qq2}) the following estimates
\begin{equation}
\vartheta^2\sim v_*L^\gamma
\frac{nP_2}{\gamma D} \,, \qquad
|y|^2\sim\frac{n\gamma D}{L^\gamma P_2 v_*} \,.
\label{mk5} \end{equation}
Recall that $v_*$ is the value of $v$ at the end of the first stage.

Substituting Eqs. (\ref{mk1}) and (\ref{mk7}) into Eq. (\ref{nn9}) we get 
\begin{equation}
G=v\left(1-{n}/{n_c}\right) \,,
\label{mk8} \end{equation}
where $n_c$ is defined by Eq. (\ref{crit}).
Recall that the monotonic behavior of $v$ implies $G>0$ and therefore 
the expression in (\ref{mk8}) is correct if $n<n_c$.
It means that for $n>n_c$ the instantonic solution of 
the considered type does not exist and we should look for another
possibility. We postpone the problem up to the next subsection and
continue to analyze the monotonic $v$ regarding $n<n_c$.

Substituting (\ref{mk7}) into Eq. (\ref{aaf}) we get
\begin{equation} 
\ln\frac{1}{v_*}=\left(1-n/n_c\right)
\gamma\ln\frac{L}{r} \,.
\label{mk9} \end{equation}
Next, we should find the leading contribution to the action. This
contribution is made mainly by the first stage producing a large
logarithm. Therefore, we can write using (\ref{nn4})
$$ i{\cal I}\approx \gamma^{-1}
\int\limits_{v_*}^{1}{\rm d}v\,\left(\mu+\frac{H_1}{G}\right) \,. $$
Substituting here Eqs. (\ref{mk6},\ref{mk7},\ref{mk8}) we find with
the same logarithmic accuracy
\begin{equation}
i{\cal I}=-\frac{(2-\gamma)n^2}{2{d}}\ln\frac{L}{r} \,.
\label{mk10} \end{equation}

Finally we can determine the structure functions $S_n$ (\ref{str})
in accordance with formula (\ref{basic}). Collecting Eqs.
(\ref{mk5},\ref{mk9},\ref{mk10})
we obtain the expressions (\ref{mk11},\ref{mk12}) of the main text.

\subsection{Remote tail}
\label{remote}

Let us proceed to discussing the character of the instantonic solution at
$n>n_c$. We can solve the corresponding equations in two limiting cases: 
$n\gg n_c$ and $n-n_c\ll n_c$. The latter case is considered in Appendix 
\ref{critic}. In the subsection we accept $n\gg n_c$, the inequality 
ensures also $v_*\gg1$ (recall that $v_*$ is the value of $v$ 
at the end of the first stage). Due to the condition $v_*\gg1$, the 
last substage of the first stage and the second stage can be examined 
in terms of the Hamiltonian
\begin{equation} 
H=-\mu v +\frac{\mu^2}{2{d}}v
+\frac{|y|^2}{D}R_+^\gamma
\left[\chi(R_+)-\chi(R_+ v^{1/\gamma})\right] \,,
\label{kn6} \end{equation}
which follows from Eqs. (\ref{nn3},\ref{ham},\ref{phi}) at $v\gg1$.
At the end of the first stage $R_-\sim L$ whereas
$R_+\approx R_-/v_*^{1/\gamma}\ll L$. The second Eq. (\ref{nn7})
shows that $\mu$ which is equal to zero at $R_->L$  varies essentially at
$R_-\sim L$. To find the value of $\mu$ at $R_-<L$, we will use
the relation
\begin{equation}
\frac{\rm d}{{\rm d}\xi}\left[(H+\mu v) R_+^{-\gamma}\right]
=\frac{|y|^2}{D}\chi'(R_+) R_+ \,,
\label{knk} \end{equation}
that follows from Eqs. (\ref{nn7},\ref{kn6}).
Actually (\ref{knk}) is the equation for $m_+$ that can be obtained from
Eqs. (\ref{lagr},\ref{qq1},\ref{nn1}) under the same conditions that led to
Eq. (\ref{kn6}). Since $R_+\ll R_-\sim L$ during the last substage of the first
stage and the second stage, the right-hand side of Eq. (\ref{knk}) 
can be neglected and we get a conservation law for $(H+\mu v) R_+^{-\gamma}$.
Equating the values of that quantity at $R_-<L$ and at $R_->L$ we find that 
at the end of the first stage
\begin{equation} 
\mu_* v_*\sim \sqrt{\frac{{d}|y|^2 P_2}{D}L^\gamma}
\label{kn7} \end{equation}

Now, let us return to relation (\ref{qq3}). Since 
$R_-\sim L \gg R_+$ at the end of the first stage the main contribution
to the integral (\ref{qq3}) is made when already $R_->L$ and $R_+$
increases to $L$. Using Eqs. (\ref{qq3},\ref{nn2}), and (\ref{qq2}) we get
\begin{equation}
\vartheta^2\sim n\frac{P_2}{\gamma D}L^\gamma \,. \qquad
|y|^2\sim n\frac{\gamma D}{P_2 L^\gamma} \,.
\label{kn8} \end{equation}
Substituting the expression (\ref{kn8})
for $|y|^2$ into (\ref{kn7}) we conclude that $\mu_* v_*$  with $n$.
In addition, at $R_-<L$ the Hamiltonian (\ref{kn6}) should be equal to 
$H_1\approx H_c$, (\ref{kn1}), that is
$$ -\mu_* v_* +\frac{\mu^2}{2{d}}v_* \approx H_c \,. $$
At $v_*\gg1$ the relation leads to
\begin{equation}
\mu_*\approx 2{d} \, \qquad
v_*\sim \sqrt{\gamma n/{d}} \,.
\label{kn9} \end{equation}
Actually, $\mu\approx 2{d}$ during the whole last substage where $v\gg1$.
Deriving the expression (\ref{kn9}) for $v_*$, we used the estimates
(\ref{kn7},\ref{kn8}).

Now, we can turn to the calculation of the effective action (\ref{nn4}).
As before, the main contribution to the action is made
by the first stage. Hence
\begin{equation}
i{\cal I}=-\frac{1}{\gamma}\int \left(
\mu\, {\rm d}v+\frac{H_1}{G}\,{\rm d}v\right)  \,. 
\label{jkn10} \end{equation}
Moreover, only the contribution related to the vicinity of the reverse point 
is relevant. Using relations (\ref{log}) we obtain
\begin{equation}
i{\cal I}\approx H_c\ln\frac{L}{r} \,.
\label{kn10} \end{equation} 
Here, we substituted $H_1\approx H_c$ and took into account that $\mu$
has no singular denominator which is clear from the relation (\ref{nn8}).
And finally, we can determine the structure functions $S_n$ (\ref{str})
in accordance with (\ref{basic}). Collecting (\ref{kn8},\ref{kn10})
we obtain the expression (\ref{kkn11}) of the main text.

Expression (\ref{kn9}) shows that indeed $v_*\gg 1$ (which is the applicability
condition of the above consideration) at $n\gg n_c$ if there are no
additional small parameters. The next subsection is devoted to special
situations appearing in the presence of such parameters.

\subsection{Special cases}
\label{special}

In the subsection we treat three special cases that are realized at $n>n_c$ if
$\gamma\ll1$, $2-\gamma\ll1$, or $n$ is extremely large. The general picture
given in Appendix \ref{remote} does not essentially change, but particular
answers should be slightly corrected.

The case $\gamma\ll1$ needs a special consideration since as follows from
Eq. (\ref{kn9}), the condition $v_*\gg1$ is violated at 
$n\sim{d}/\gamma\gg n_c\sim{d}\gamma$. Therefore, an
intermediate region exists at $n_c\ll n\ll{d}/\gamma$
where $\gamma\ll v_* \ll 1$. Then we get from Eq. (\ref{ham})
\begin{equation}
H=-\mu v +\frac{\mu^2}{2{d}}
+\frac{|y|^2}{D}R_+^\gamma
\left[\chi(R_+)-\chi(R_-)\right] \,, 
\label{kn11} \end{equation}
where 
\begin{equation}
R_-=R_+\exp\left({v}/{\gamma}\right) \,.
\label{kjn11} \end{equation}
Equations (\ref{nn7}) now read
\begin{eqnarray} &&
\frac{{\rm d}v}{{\rm d}\xi}=-\gamma v +\gamma\frac{\mu}{d} \,,
\label{kn12} \\ &&
\frac{{\rm d}\mu}{{\rm d}\xi}=\gamma\mu
+\frac{|y|^2}{D}R_+^\gamma R_-\chi'(R_-) \,.
\label{kn13} \end{eqnarray}
At the end of the first stage $R_-\sim L$. As follows from relation
(\ref{kjn11}), due to $v_*\gg\gamma$ the estimate $R_+\ll L$ is valid.
Therefore, the integral (\ref{qq3}) for $\vartheta^2$ is determined by the time
interval between the moments when $R_+$ and $R_-$ go through $L$. We can again
pass to integrating over $R_+$ in accordance with Eq. (\ref{nn2}). Though
$R_+\ll L$ at $R_-\sim L$, 
$\gamma$$R_+^\gamma\approx R_-^\gamma(1-v)$ due
to smallness of $\gamma$. Therefore, we return to the estimates (\ref{mk5}).
Next we can estimate $\mu_*$ from Eq. (\ref{kn13}) where the term $\gamma\mu$
is dropped. Since $R_+^\gamma\sim L^\gamma$ at $R_-\sim L$,
integrating over $\xi$ we obtain from Eq. (\ref{kn13})
\begin{equation}
\mu_*\sim\frac{|y|^2}{D}L^\gamma P_2 \,.
\label{kn14} \end{equation}
To justify the estimate (\ref{kn14}), one should check that
$v$ does not vary essentially in the region, where $R_-\sim L$. 
Using Eq. (\ref{kn12}) we obtain that the condition is
satisfied. Substituting formulas (\ref{mk5}) into (\ref{kn14}) we obtain 
the previous relation $\mu_* v_*\sim\gamma n\gg |H_c|$. Equating then the 
Hamiltonian (\ref{kn11}) to $H_c$ at $v=v_*$ we get
\begin{equation}
\mu_*\sim\sqrt{{d}\gamma n} \,, \qquad
v_*\sim\sqrt{\frac{\gamma n}{d}} \,.
\label{kn15} \end{equation}
The estimate (\ref{kn15}) for $v_*$ surprisingly coincides with (\ref{kn9}).
Next, we obtain from (\ref{mk5})
\begin{equation}
\vartheta^2\sim\sqrt\frac{n}{n_c}L^\gamma\frac{n P_2}{D} \,.
\label{kn16} \end{equation}
Therefore, the expression (\ref{kkn11}) should be corrected and we get
\begin{eqnarray} &&
S_n\sim\left(\sqrt\frac{n}{n_c}\frac{P_2 C_3}{D}
L^\gamma\right)^{n/2}\left(\frac{r}{L}\right)^{\zeta_c} \,,
\label{awg} \end{eqnarray}
where $C_3$ is a non-universal parameter of order unity and $\zeta_c$
is defined by Eq. (\ref{zc}).

Some peculiarities appear also at $2-\gamma\ll1$. In this case 
the function (\ref{phi}) can be approximated as 
$$ \phi\approx 2{d} v\ln v \,, $$
if $v\gg1$ but $(2-\gamma)\ln v\ll1$. Then we obtain from Eq. (\ref{ham}) 
\begin{eqnarray}&&
H=-\mu v+\frac{2-\gamma}{2{d}}v\ln v \mu^2
\nonumber\\&&+\frac{|y|^2}{D}R_+^\gamma
\left[\chi(R_+)-\chi(R_+v^{1/\gamma})\right] \,.
\label{kn17} \end{eqnarray}
Expression (\ref{kn17}) has the same structure as (\ref{kn6})
except for the logarithmic factor. Taking into account $v\gg1$
we conclude that the Hamiltonian (\ref{kn17}) leads to the conservation
law (\ref{knk}) which is satisfied with the same accuracy $1/v$ as
previously. Therefore, we get instead of (\ref{kn7})
\begin{equation} 
\mu_* v_*\sim \sqrt{\frac{{d}|y|^2 P_2}
{D(2-\gamma)\ln v_*}L^\gamma} \sim
\sqrt{\frac{n{d}}{(2-\gamma)\ln v_*}} \,,
\label{kn18} \end{equation}
where we used (\ref{kn8}), its validity is accounted for the inequality
$v\gg1$. One can check that the main contribution to
$H$ in (\ref{kn17}) at the end of the first stage is made by $-\mu v$.
Equating $-\mu_* v_*$ to $H_c$ we get
\begin{equation}
\ln v_*\sim \frac{n}{n_c} \,.
\label{kn19} \end{equation}
Now we obtain the condition $n\ll n_c/(2-\gamma)$, under which the
regime under consideration is realized. Since $v\gg 1$ in the regime we
have the same expression (\ref{kn8}) for $\vartheta$ and, consequently,
the same estimate (\ref{kkn11}) for the structure functions.

We finish this subsection with a discussion concerning extremely large $n$.
In accordance with the estimates (\ref{kn9}) $v_*$ increases with increasing
$n$.
Therefore, the last substage of the first stage where $v\gg1$ starts
to play an essential role. First of all we should correct Eq. (\ref{log})
returning to Eq. (\ref{fv0}). If  $v=v_*$ there then $v_* R_+^\gamma$ in 
the right-hand side of the relation can be substituted by $L^\gamma$.
The contribution to the integral in the left-hand side of Eq. (\ref{log})
associated with the last substage (when $v$ increases from $1$ to $v_*$)
can be found substituting $G(x)\approx-x$ there. Thus the substage 
makes the additional logarithmic contribution to the integral. 
Taking the contribution into account we get instead of Eq. (\ref{kn5})
\begin{equation}
\sqrt{\frac{{d}}{8(2-\gamma)}}\frac{2\pi}{\sqrt{(H_c-H_1)}} 
=\gamma\ln\frac{L}{r}-2\ln{v_*} \,.
\label{kmn5} \end{equation}
There appears also an analogous correction to expression (\ref{kn10}).
Calculating the action (\ref{jkn10}) we get
\begin{equation}
i{\cal I}\approx H_c\left(\ln\frac{L}{r}
-2\frac{\ln{v_*}}{\gamma}\right) \,.
\label{Cwg} \end{equation}
Here we neglected the term originated from $\int\mu\,{\rm d}v$ since
using (\ref{kn9}) the term can be estimated as $\sqrt{n/(\gamma{d})}$ 
and is consequently negligible in comparison with the terms $\sim n$
entering $S_n$ via $\vartheta$. Substituting the expression (\ref{Cwg})
into (\ref{basic}) we conclude that the correction related to the last
contribution in (\ref{Cwg}) can be neglected in comparison to the
strong $n$-dependence of $\vartheta^n$. Nevertheless, the presence of the
additional term in the right-hand side of Eq. (\ref{kmn5}) shows that
our approach is broken at $2\ln v_*\sim \gamma \ln({L}/{r})$ since
$H_c-H_1$ ceases to be a small there. Using the estimates (\ref{kn9}) we
arrive at the condition (\ref{jkn12}), which is the 
applicability condition of our approach. Note that at 
$\ln({n}/{d})\gtrsim\gamma\ln({L}/{r})$ a scaling behavior 
of $S_n$ is destroyed.

\subsection{Instanton near critical order}
\label{critic}

Here we will consider the case when $n$ is close to the critical value
(\ref{crit}): $|n-n_c|\ll n_c$. Then $v_*\ll1$ and at examining the last
substage of the first stage and of the second stage we can use the Hamiltonian,
expanded over small $v$.

If $n<n_c$ then we can take the Hamiltonian (\ref{mk2}). The only peculiarity 
of the consideration near $n_c$ is that the logarithmic contributions to the
effective action (\ref{nn4}) and to the left-hand side of the integral
(\ref{fv0}) depend on $n-n_c$. Indeed, the expression (\ref{kn2}) shows that
$G(x)$ is a linear function of $x$ only in the restricted interval
$$ \frac{2-\gamma}{2{d}\gamma}|H_c-H_1|> v>v_* \,. $$ 
Just the interval determines the value of the logarithm. It
means that we should substitute in the expressions (\ref{mk9},\ref{mk10}) 
$$ \ln\frac{1}{v_*}\to\ln\frac{v_* n_c^2}{(n-n_c)^2} \,. $$
The substitution does not change the final answer (\ref{mk10}) for the action
but influences the value of $\vartheta^2$ because of (\ref{mk5}). Taking the
fact into account we get the expression (\ref{strcr}) that replace (\ref{mk11}).

It is clear that at $n\to n_c$ the difference $n-n_c$ starts to compete with
$r/L$ what destroys our construction. Let us estimate the corresponding value
of $n_c-n$ regarding $n<n_c$. It follows from the asymptote (\ref{kn2}) that
our consideration is correct if $(H_c-H_1)\gg{d}\gamma v_*/(2-\gamma)$. 
Expressing in the inequality $v_*$ via $r/L$ from Eq. (\ref{aaf}) we obtain the
criterion (\ref{crit1}) written in the main text. Note that the criterion
implies the inequality
$$ \gamma\ln\frac{L}{r}\gg1 \,,  $$
which should be assumed for our treatment to be correct. Otherwise a special 
consideration is needed.

Let us now proceed to the case $n>n_c$. We should take into account
the next term of the expansion of $H$ over $v$ in comparison with 
(\ref{mk2}) which should correct the value (\ref{mk7}) of $H_1$ since 
we know that $H_1$ is equal to $H_c$. The corresponding expression is 
\begin{eqnarray}&&
H=-\mu v+\frac{2(2-\gamma)}{{d}\gamma^2}
\left(1-\frac{4-\gamma}{2\gamma}v\right)\mu^2 v^2\nonumber\\&&
+\frac{|y|^2}{D}R_+^\gamma
\left[\chi(R_+)-\chi(R_-)\right] \,,
\label{jk1} \end{eqnarray}
where the difference $\chi(R_+)-\chi(R_-)$ should be expanded up to the second 
order over $v$. Eqs. (\ref{nn7}) with the Hamiltonian (\ref{jk1}) lead to
\begin{eqnarray} &&
\frac{1}{\gamma}\frac{{\rm d}v}{{\rm d}\xi}
=-v+\mu v^2(\alpha_1-\alpha_2 v) \,,
\label{jk3} \\ &&
\frac{1}{\gamma}\frac{\rm d}{{\rm d}\xi}(\mu v)
=\frac{1}{2}\alpha_2 \mu^2 v^3
+\frac{|y|^2}{D}R_+^{\gamma+1}\chi'(R_-) v \,,
\label{jk4} \\ &&
\alpha_1=\frac{4(2-\gamma)}{{d}\gamma^2}\,, \qquad
\alpha_2=\frac{2(4-\gamma)(2-\gamma)}{{d}\gamma^2}\,. 
\label{jk12}  \end{eqnarray} 
Equating the Hamiltonian (\ref{jk1}) to the value (\ref{kn1}) we get
\begin{equation}
\mu v\approx \frac{\gamma n_c}{2} \,,
\label{lf11} \end{equation}
at the end of the first stage. Note that due to $v_*\ll1$ we can use the 
estimates (\ref{mk5}). 

Our next aim is to estimate $v_*$ in terms of $n-n_c$. To do it,
we will use the identity $dH/{\rm d}\xi=\partial H/\partial\xi$ that is correct
for any canonical system. Let us integrate the relation over the second 
stage. Since $H=H_1$ at the first stage and $H=0$ at the end of the second 
stage we obtain
\begin{eqnarray}&&
-H_1=\frac{|y|^2}{D}\int {\rm d}\xi\, 
\Bigl\{\gamma R_+^\gamma\left[\chi(R_+)-\chi(R_-)\right]
\nonumber \\&&+ R_+^\gamma\left[R_+\chi'(R_+)-R_-\chi'(R_-)\right]\Bigr\} \,.
\label{jk2} \end{eqnarray} 
Expanding the right-hand side of relation (\ref{jk2}) up to
the second order over $v$, integrating by parts (to remove high
derivatives of $\chi$) and using Eqs.
(\ref{nnk},\ref{jk3},\ref{jk4},\ref{qq2}) one can obtain the
expression
\end{multicols}
\begin{eqnarray} &&
-H_1=\frac{\gamma n}{2}
\left(1-\frac{\gamma\phi_1 n}{4}\right)
-\frac{\gamma\alpha_1^2}{2}(n-n_c)\frac{|y|^2}{D}
\int {\rm d}\xi\,R_+^{1+\gamma}\chi'(R_+)\mu v^3
\nonumber \\ &&
-\frac{|y|^2\alpha_2}{D}
\int {\rm d}\xi\,R_+^{1+\gamma}\chi'(R_+)\mu v^3
-\frac{\gamma|y|^2}{2D}\alpha_1\alpha_2
\int {\rm d}\xi\,R_+^{1+\gamma}\chi'(R_+) v
\int\limits_\xi^\infty{\rm d}\tilde\xi\,\mu^2 v^3 \,,
\label{jk11} \end{eqnarray}
\begin{multicols}{2}

\noindent
The first term in the right-hand side of (\ref{jk11}) reproduces the
previous result (\ref{mk7}) and other terms are small corrections 
proportional to $v_*$. 

We know that in the main approximation over $r/L$ we can substitute $H_1=H_c$. 
Regarding $n-n_c\ll n_c$ we neglect the second term in the right-hand side of
Eq. (\ref{jk11}). The last two terms in the right-hand side of Eq.
(\ref{jk11}) can be estimated using
$\mu_* v_*\sim \gamma n_c$, $\chi\sim P_2$, $R_+\sim L$ and the
estimation (\ref{mk5}) for $|y|^2$. Combining all together we get
\begin{equation} 
v_*\sim\gamma\frac{(n-n_c)^2}{n_c^2} \,.
\label{jk14} \end{equation}
Let us stress that due to $\alpha_2>0$ the last two terms in the right-hand
side of (\ref{jk11}) are positive if $\chi(R)$ is a monotonic function which is
a reasonable condition. Therefore, a solution of Eq. (\ref{jk11}) for
$v_*$, determined by the estimate (\ref{jk14}), really exists. We conclude
that there is an instantonic solution with non-monotonic behavior of $v$ at any
$n-n_c\ll n_c$.

The last assertion should be corrected since it is true only if $L/r\to\infty$.
At finite $L/r$ there is a narrow region of very small $n-n_c$ where our scheme
does not work. To establish the corresponding criterion let us remind that our
consideration is valid if $v_r\ll v_*$.  It follows from Eq. (\ref{kn5}) that
$v_r\sim 1/[\gamma\ln^2(L/r)]$. Then, from formula (\ref{jk14}) we get the
inequality (\ref{crit1}).

And finally, we can determine the structure functions $S_n$ (\ref{str}) in
accordance with (\ref{basic}). Collecting Eqs. 
(\ref{mk5},\ref{kn10},\ref{jk14}) we obtain the formula (\ref{strcr}) of the
main text.

\end{multicols}


\begin{references}

\bibitem{Kol}
A. N. Kolmogorov, C. R. Acad. Sci. URSS, {\bf 30}, 301 (1941).
\bibitem{61GSM}
H. L. Grant, R. W. Stewart, and A. Moilliet,
J. Fluid Mech. {\bf 12}, 241 (1961).
\bibitem{62BT}
G. K. Batchelor and A. A. Townsend,
Proc. R. Soc. London Sect. A {\bf 199}, 82 (1962).
\bibitem{91MS}
C. Meneveau and K. R. Sreenivasan,
J. Fluid Mech. {\bf 224}, 429 (1991).
\bibitem{53Bat}
G. K. Batchelor, {\it Theory of Homogeneous Turbulence}
(Cambridge University Press, New York, 1953).
\bibitem{MY}
A. Monin and A. Yaglom, {\it Statistical Fluid Mechanics}
(MIT Press, Cambridge, 1975).
\bibitem{Frish}
U. Frisch, {\it Turbulence: the Legacy of A. N. Kolmogorov}
(Cambridge University Press, New York, 1995).  
\bibitem{Obukh}
A. M. Obukhov, C. R. Acad. Sci. URSS, Geogr. Geophys., {\bf 13}, 58 (1949).
\bibitem{Corrs}
S.~Corrsin, J. Appl. Phys., {\bf 22}, 469, (1951).
\bibitem{84AHGA}
R. Antonia, E. Hopfinger, Y. Gagne, and F. Anselmet, 
Phys.~Rev.~A {\bf 30}, 2704 (1984).
\bibitem{90MS}
C. Meneveau and K.~R. Sreenivasan, Phys.~Rev.~A {\bf 41}, 2246 (1990).
\bibitem{91Sre}
K.~R. Sreenivasan, Proc. R. Soc. (London) A {\bf 434}, 165 (1991).
\bibitem{91HY}
I.~Hosokawa and K.~Yamamoto, in {\em Turbulence and Coherent Structures},
ed. by O.~Metais and M.~Lesieur (Kluwer, London 1991).
\bibitem{68Kra-a}
R.~H. Kraichnan, Phys.~Fluids {\bf 11}, 945 (1968).
\bibitem{95SS}
B.~Shraiman and E.~Siggia,
C. R. Acad. Sc. {\bf 321}, Ser. II, 279 (1995).
\bibitem{95GK}
K.~Gawedzki and A.~Kupiainen,
Phys.~Rev.~Lett. {\bf75}, 3608 (1995).
\bibitem{95CFKLb}
M. Chertkov, G. Falkovich, I. Kolokolov and V. Lebedev,
Phys. Rev. E {\bf 52}, 4924 (1995).
\bibitem{95CF}
M.~Chertkov and G.~Falkovich,
Phys.~Rev.~Lett. {\bf 76}, 2706 (1995).
\bibitem{96BGK}
D.~Bernard, K.~Gawedzki and A.~Kupiainen, 
Phys. Rev. E {\bf 54}, 2564 (1996).
\bibitem{96SS}
B.~Shraiman and E.~Siggia,
Phys.~Rev.~Lett. {\bf 77}, 2463 (1996).
\bibitem{96BG}
E. Balkovsky and D. Gutman, 
Phys. Rev. E {\bf 54}, 4435 (1996).
\bibitem{97BFL}
E. Balkovsky, G. Falkovich, and V. Lebedev,
Phys. Rev. E {\bf 55}, R4881, (1997).
\bibitem{95KYC}
R. H. Kraichnan, V. Yakhot, and S. Chen, 
Phys. Rev. Lett. {\bf 75}, 240 (1995).
\bibitem{97FGLP}
A.~L.~Fairhall, B.~Galanti, V.~S.~L'vov, and I.~Procaccia, 
Phys. Rev. Lett. {\bf 79}, 4166, (1997).
\bibitem{98FMV}
U.~Frisch, A.~Mazzino, and M.~Vergassola,
submitted to Phys. Rev. Lett. (1998).
\bibitem{closure}
R. H. Kraichnan, Phys. Rev. Lett. {\bf 52}, 1016 (1994). 
\bibitem{97Yak} 
V.~Yakhot, Phys. Rev. E {\bf 55}, 329 (1997).
\bibitem{97Chert} M. Chertkov,
Phys. Rev. E {\bf 55}, 2722 (1997).
\bibitem{73MSR}
P.~C. Martin, E.~Siggia, and H.~Rose,
Phys.~Rev.~A{\bf 8}, 423 (1973).
\bibitem{76Dom}
C.~de~Dominicis, J.~Physique (Paris) {\bf 37}, c01-247 (1976).
\bibitem{76Jan}
H. Janssen, Z.~Phys.~B {\bf 23}, 377 (1976).
\bibitem{98BL}
E. Balkovsky and V. Lebedev, submitted to Phys. Rev. Lett.
\bibitem{96FKLM}
G.~Falkovich, I.~Kolokolov, V.~Lebedev and A.~Migdal,
Phys. Rev. E{\bf 54}, 4896 (1996).
\bibitem{96GM}
V.~Gurarie and A.~Migdal,
Phys. Rev. E{\bf 54}, 4908 (1996).
\bibitem{97BFKL}
E.~Balkovsky, G.~Falkovich, I.~Kolokolov and V.~Lebedev,
Phys. Rev. Lett. {\bf 78}, 1452 (1997).
\bibitem{97FL}
G.~Falkovich and V.~Lebedev,
Phys. Rev. Lett. {\bf 79}, 4159 (1997).
\bibitem{98BF}
E.~Balkovsky and G.~Falkovich,
Phys. Rev. E {\bf 57}, 1231 (1998).
\bibitem{94SS}
B.~Shraiman and E.~Siggia,
Phys. Rev. E {\bf 49}, 2912 (1994).
\bibitem{26Rich}
L. Richardson, Proc. Roy. Soc. 
(London) A {\bf 110}, 709 (1926).
\bibitem{62Kol}
A. N. Kolmogorov, J. Fluid Mech., {\bf 12}, 82 (1962).
\bibitem{Les}
M.~Lesieur, {\it Turbulence in Fluids}
(Kluwer Academic Publishers, Dordrecht, 1997).
\bibitem{98BGK}
D.~Bernard, K.~Gawedzki and A.~Kupiainen, 
J. Stat. Phys. {\bf 90}, 519 (1998).
\bibitem{98FMVa}
U.~Frisch, A.~Mazzino, and M.~Vergassola, in
Proceedings of Europ. Geophys. Soc. meeting, 1998.
\bibitem{66Kra-b} R.~H. Kraichnan,
Phys. Fluids, {\bf9}, 1937 (1966).

\end{references}
\end{document}